\documentclass[twocolumn]{aastex62}
\usepackage{graphicx}
\usepackage{amsmath}

\graphicspath{{./}{figures/}}

\accepted{October 7, 2019}
\submitjournal{ApJ}

\shorttitle{SNe~II in the NIR}
\shortauthors{Davis et al.}

\begin{document}

\title{\textit{Carnegie Supernova Project-II:} Near-infrared Spectroscopic Diversity of Type II Supernovae\footnote{This paper includes data gathered with the 6.5 meter \textit{Magellan} Telescopes at Las Campanas Observatory, Chile.}}

\correspondingauthor{Scott Davis}
\email{sfd15b@my.fsu.edu}

\author[0000-0002-2806-5821]{S. Davis}
\affil{Department of Physics, Florida State University, 77 Chieftan Way, Tallahassee, FL 32306, USA}
\author[0000-0003-1039-2928]{E. Y. Hsiao}
\affil{Department of Physics, Florida State University, 77 Chieftan Way, Tallahassee, FL 32306, USA}
\author[0000-0002-5221-7557]{C. Ashall}
\affil{Department of Physics, Florida State University, 77 Chieftan Way, Tallahassee, FL 32306, USA}
\author[0000-0002-4338-6586]{P. Hoeflich}
\affil{Department of Physics, Florida State University, 77 Chieftan Way, Tallahassee, FL 32306, USA}
\author[0000-0003-2734-0796]{M. M. Phillips}
\affiliation{Carnegie Observatories, Las Campanas Observatory, Casilla 601, La Serena, Chile}
\affiliation{Wyatt-Green Preeminent Visiting Scholar of the Department of Physics, Florida State University}

\author{G.~H.~Marion}
\affiliation{University of Texas at Austin, 1 University Station C1400, Austin, TX, 78712-0259, USA}
\author[0000-0002-1966-3942]{R. P. Kirshner}
\affiliation{Gordon and Betty Moore Foundation, 1661 Page Mill Road, Palo Alto, CA 94304 }
\affiliation{Harvard-Smithsonian Center for Astrophysics, 60 Garden Street, Cambridge, MA 02138}
\author[0000-0003-2535-3091]{N. Morrell}
\affiliation{Carnegie Observatories, Las Campanas Observatory, Casilla 601, La Serena, Chile}
\author[0000-0003-4102-380X]{D. J. Sand}
\affil{Department of Astronomy/Steward Observatory, 933 North Cherry Avenue, Rm. N204, Tucson, AZ 85721-0065, USA}
\author[0000-0003-4625-6629]{C. Burns}
\affiliation{Observatories of the Carnegie Institution for Science, 813 Santa Barbara St., Pasadena, CA 91101, USA }
\author[0000-0001-6293-9062]{C. Contreras }
\affiliation{Carnegie Observatories, Las Campanas Observatory, Casilla 601, La Serena, Chile}
\author[0000-0002-5571-1833]{M. Stritzinger }
\affiliation{Department of Physics and Astronomy, Aarhus University, 
Ny Munkegade 120, DK-8000 Aarhus C, Denmark }

\author[0000-0003-0227-3451]{J. P. Anderson}
\affil{European Southern Observatory, Alonso de C\'ordova 3107, Casilla 19, Santiago, Chile}

\author[0000-0001-5393-1608]{E. Baron}
\affiliation{Department of Physics and Astronomy, Aarhus University, Ny Munkegade 120, DK-8000 Aarhus C, Denmark }
\affiliation{Homer L. Dodge Department of Physics and Astronomy, University of Oklahoma, 440 W. Brooks, Rm 100, Norman, OK 73019-2061, USA}
\affiliation{Hamburger Sternwarte, Gojenbergsweg 112, D-21029 Hamburg, Germany}

\author[0000-0002-0805-1908]{T. Diamond}
\affiliation{NASA Goddard Space Flight Center, Greenbelt, MD 20771, USA}

\author{C. P. Guti\'{e}rrez}
\affiliation{Department of Physics and Astronomy, University of Southampton,
Southampton, SO17 1BJ, UK}

\author[0000-0001-7981-8320]{M. Hamuy}
\affil{Departamento de Astronom\'{i}a, Universidad de Chile, Casilla 36-D, Santiago, Chile}

\author{S. Holmbo}
\affiliation{Department of Physics and Astronomy, Aarhus University, 
Ny Munkegade 120, DK-8000 Aarhus C, Denmark }

\author[0000-0002-5619-4938]{M. M. Kasliwal}
\affil{4Division of Physics, Mathematics and Astronomy, California Institute of Technology, Pasadena, California 91125, USA}

\author[0000-0002-6650-694X]{K. Krisciunas }
\affiliation{George P. and Cynthia Woods Mitchell Institute for Fundamental Physics \& Astronomy, Texas A\&M University, Department of Physics, 4242 TAMU, College Station, TX 77843}

\author[0000-0001-8367-7591]{S. Kumar}
\affil{Department of Physics, Florida State University, 77 Chieftan Way, Tallahassee, FL 32306, USA}

\author[0000-0002-3900-1452]{J. Lu}
\affil{Department of Physics, Florida State University, 77 Chieftan Way, Tallahassee, FL 32306, USA}

\author[0000-0002-8041-8559]{P.J. Pessi}
\affil{Instituto de Astrof\'{i}sica de La Plata (IALP), CONICET, Paseo del bosque S/N, 1900, Argentina}
\affil{Facultad de Ciencias Astron\'{o}micas y Geof\'{i}sicas (FCAG), Universidad Nacional de La Plata (UNLP), Paseo del bosque S/N, 1900, Argentina.}

\author[0000-0001-6806-0673]{A. L. Piro}
\affil{Observatories of the Carnegie Institution for Science, 813 Santa Barbara St, Pasadena, CA 91101, USA}

\author{J. L. Prieto}
\affiliation{N\'{u}cleo de Astronom\'{i}a, Facultad de Ingenier\'{i}a y Ciencias, Universidad Diego Portales, Ej\'{e}rcito 441, Santiago, Chile}

\author[0000-0002-9301-5302]{M. Shahbandeh}
\affiliation{Department of Physics, Florida State University, 77 Chieftan Way, Tallahassee, FL 32306, USA}

\author[0000-0002-8102-181X]{N. B. Suntzeff}
\affiliation{George P. and Cynthia Woods Mitchell Institute for Fundamental Physics \& Astronomy, Texas A\&M University, Department of Physics, 4242 TAMU, College Station, TX 77843}

\begin{abstract}

We present $81$ near-infrared (NIR) spectra of $30$ Type II supernovae (SNe~II) from the \textit{Carnegie Supernova Project-II} (CSP-II), the largest such dataset published to date. We identify a number of NIR features and characterize their evolution over time. The NIR spectroscopic properties of SNe II fall into two distinct groups. This classification is first based on the strength of the He\,{\footnotesize I} $\lambda1.083\,\mu$m absorption during the plateau phase; SNe~II are either significantly above (spectroscopically \emph{strong}) or below $50$ \AA{} (spectroscopically \emph{weak}) in pseudo equivalent width. However between the two groups, other properties, such as the timing of CO formation and the presence of Sr\,{\footnotesize II}, are also observed. Most surprisingly, the distinct \emph{weak} and \emph{strong} NIR spectroscopic classes correspond to SNe~II with slow and fast declining light curves, respectively. These two photometric groups match the modern nomenclature of SNe~IIP and IIL. Including NIR spectra previously published, 18 out of 19 SNe~II follow this slow declining-spectroscopically \emph{weak} and fast declining-spectroscopically \emph{strong} correspondence. This is in apparent contradiction to the recent findings in the optical that slow and fast decliners show a continuous distribution of properties. 
The \emph{weak} SNe~II show a high-velocity component of helium that may be caused by a thermal excitation from a reverse-shock created by the outer ejecta interacting with the red supergiant wind, but the origin of the observed dichotomy is not understood. Further studies are crucial in determining whether the apparent differences in the NIR are due to distinct physical processes or a gap in the current data set.

\end{abstract}

\section{Introduction} \label{intro}
Type II supernovae (SNe~II) are classified by the presence of Balmer-series hydrogen lines in their optical spectra \citep{1941PASP...53..224M}. They are believed to result from the explosion of massive stars ($>8M_{\odot}$) that have retained a significant portion of their hydrogen envelope. Pre-explosion images of SN~II locations suggest that the majority of SNe~II are the result of red supergiant (RSG) explosions \citep[e.g.][]{2004Sci...303..499S,2005MNRAS.364L..33M,2009ARA&A..47...63S,2015PASA...32...16S,2017hsn..book..693V,2019ApJ...875..136V}.

Historically, SNe~II have been divided into subclasses based on the shape of their light curves \citep{1979A&A....72..287B}. In this work the more modern nomenclature for SNe~II photometric classifications is used, with SNe exhibiting fast declining light curves classified as SNe~IIL, and those with slow declining light curves classified as SNe~IIP. 
The majority of SN~II $V$-band light curves evolve with an initial rise to maximum followed by a steep decline before settling onto a more gentle decline, commonly referred to as the plateau. During this plateau phase, hydrogen recombination drives the light curve until a sharp decline onto the radioactive tail \citep{1987BAAS...19.1036W}. One photometric and two spectroscopic classes were added within the SN~II population: SNe~IIb, SNe~IIn, and SN~1987A-like, respectively \citep{1989ARA&A..27..629A,1990MNRAS.244..269S,1993ApJ...415L.103F}. 
For the remainder of this paper, we will use the SNe~II designation to refer to the collection of SNe that previously have been classified as SNe~IIL or SNe~IIP as the other previously mentioned sub-types (IIb, IIn, and SN~1987A-like) are not included in the data analyzed.

It has been suggested that fast-declining SNe~II have a smaller amount of hydrogen in their envelopes than slow-declining SNe~II \citep{1993ApJ...414..712P,2014MNRAS.445..554F,2014ApJ...786L..15G,2016MNRAS.455..423M}. However, it is still not known whether there are progenitor differences of explosion scenarios that separate the slow and fast declining SNe~II into two distinct groups \citep{1979A&A....72..287B,1993A&AS...98..443P}. Furthermore, recent publications have shown that there is no discrete photometric separation between these SNe \citep[e.g.][]{2014ApJ...786...67A,2015ApJ...799..208S,2016MNRAS.459.3939V,2016yCat..51510033G,2019MNRAS.tmp.1836P}; although see \citet{2012ApJ...756L..30A} for a possible separation. 

The amount of SN~II optical data obtained has greatly increased in the past decade with the focus turning towards larger samples to examine their spectroscopic and photometric diversity \citep{2014ApJ...786...67A,2014MNRAS.445..554F,2014ApJ...786L..15G,2016MNRAS.459.3939V,2016yCat..51510033G,2016ApJ...828..111R,2017ApJ...850...89G,2017ApJ...850...90G,2018MNRAS.473..513F}. Large photometric studies have defined parameters useful for characterizing the diversity of SN~II optical light curves. They found that brighter SNe~II decline more quickly at every phase, have shorter plateau phases, and higher ${}^{56}$Ni masses. This is significant because it further supports that faster declining SNe~II are the result of explosions with lower hydrogen envelope masses, which causes their shorter plateau duration. \cite{2017ApJ...850...89G}, using optical spectra, found that SNe~II span a continuous range in equivalent widths and velocities.

SNe~II in the NIR have yet to be explored in detail. However, there are well-known advantages to observing spectroscopically in the NIR, namely, the variety of strong lines present, less line blending when compared to the optical, and a lower optical depth revealing the core at earlier times \citep{1993MNRAS.261..535M}. Furthermore, at late times, carbon monoxide (CO) is sometimes observed \citep{1988Natur.334..327S,2001A&A...376..188S,2019AAS...23333502R}.

Spectroscopically, the most well-studied SN~II in the NIR is SN~1987A \citep{1987ESOC...26..159B,1987IAUC.4484....1O,1988MNRAS.231P..75C,1988ApJ...331L...9E,1988ApJ...325L...1R,1988MNRAS.234P...5W,1989MNRAS.237P..55C,1989MNRAS.238..193M,1990Ap&SS.171..213S,1991A&A...245..490B,1991supe.conf...69D,1991supe.conf..102M,1993MNRAS.261..535M} with over 30 NIR spectra of SN~1987A available. SN~1987A exploded in the Large Magellanic Cloud, which allowed for extensive spectroscopic and photometric coverage still being obtained today. SN~1987A was a peculiar SN~II, that most likely originated from a blue supergiant progenitor \citep{1992PASP..104..717P}. This peculiarity led to the definition of an additional SN~II subclass, SN~1987A-like, which are SNe that have long rising light curves.

Most previously published SNe~II NIR spectra are of a single SN across a small number of epochs \citep{2001MNRAS.322..361B,2001ApJ...558..615H,2003MNRAS.338..939E,2006MNRAS.368.1169P,2009MNRAS.394.2266P,2010MNRAS.404..981M,2011MNRAS.417.1417F,2013MNRAS.434.1636T,2014PASJ...66..114M,2014ApJ...787..139D,2014MNRAS.439.3694J,2014MNRAS.438..368T,2015MNRAS.450.3137T,2015MNRAS.448.2608V,2016MNRAS.459.3939V,2019AAS...23333502R,2019MNRAS.485.5120B,2019ApJ...876...19S}. These works have been able to characterize the general spectral line evolution from a couple days past explosion through the nebular phase. The SN~IIP SN~1999em was particularly well studied allowing line identification and analysis \citep{2001ApJ...558..615H,2003MNRAS.338..939E}, however, SNe~II are diverse and thus SNe~II in the NIR have not been fully explored.

By modeling the NIR and optical spectra of SNe~II, \cite{2014MNRAS.438..368T} found that the $0.98-1.12\,\mu$m region contains a high-velocity (HV) helium feature on the blue side of the Paschen gamma ($P_{\gamma}$) hydrogen absorption. This feature, previously unconfirmed, mirrors the velocity of the HV hydrogen feature claimed in the optical by \cite{2017ApJ...850...89G}.

In this paper we present the largest SNe~II NIR dataset published to date. The data were obtained between 2011 to 2015 as a part of the \textit{Carnegie Supernova Project-II} \citep[CSP-II, e.g.][]{2019PASP..131a4001P}; CSP-II was an NSF-supported program to study Type Ia SNe as distance indicators, that aimed to improve upon CSP-I \citep{2006PASP..118....2H} by observing objects from untargeted searches detected in the Hubble flow. CSP-II also worked to obtain a large sample of SN Ia NIR spectra for studies in cosmology and in their explosion physics \citep{2019PASP..131a4002H}. Furthermore, NIR spectra of all types of nearby SNe were obtained as the current sample is small and they are crucial for understanding the origins of these explosions.

The sections of this paper are outlined as follows. Section \ref{observations} is an overview of the data sample, including the observation and reduction techniques used. In Section \ref{measurements}, we describe the process used for measuring various photometric and spectroscopic properties. In Section \ref{lineID}, we outline the NIR spectral features and their evolution over time. In Section \ref{quantMeasure}, properties of the NIR hydrogen features are discussed. Section \ref{dichotomy} describes the observed NIR spectral dichotomy. In Section \ref{PCA}, we present the results of a principal component analysis (PCA) on the CSP-II SNe~II dataset and corresponding NIR spectral templates. The discussion of results and conclusions are in Sections \ref{discussion} and \ref{conclusions}, respectively.

\section{Observations and Sample} \label{observations}
\begin{figure}
    \includegraphics[width=\columnwidth]{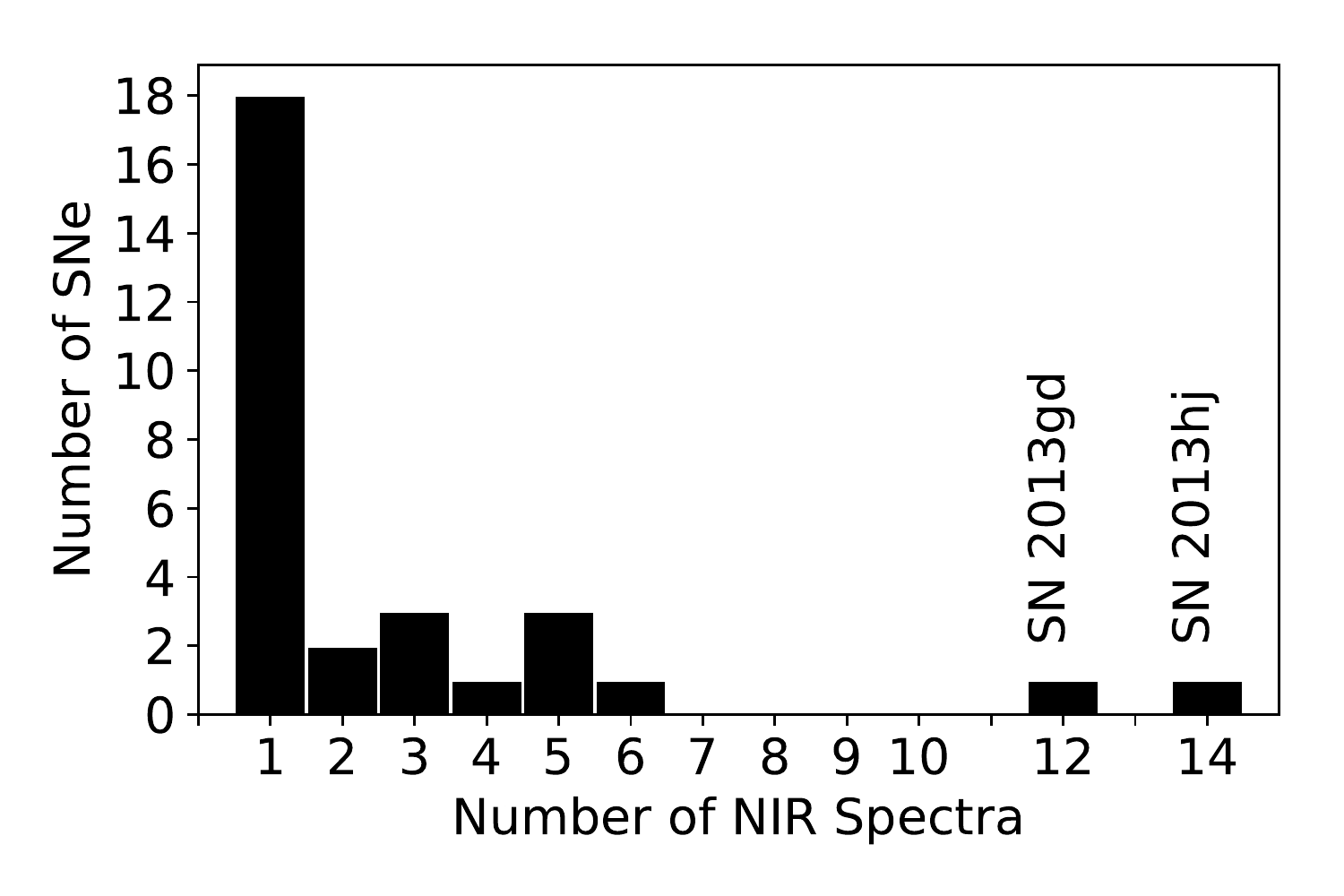}
    \includegraphics[width=\columnwidth]{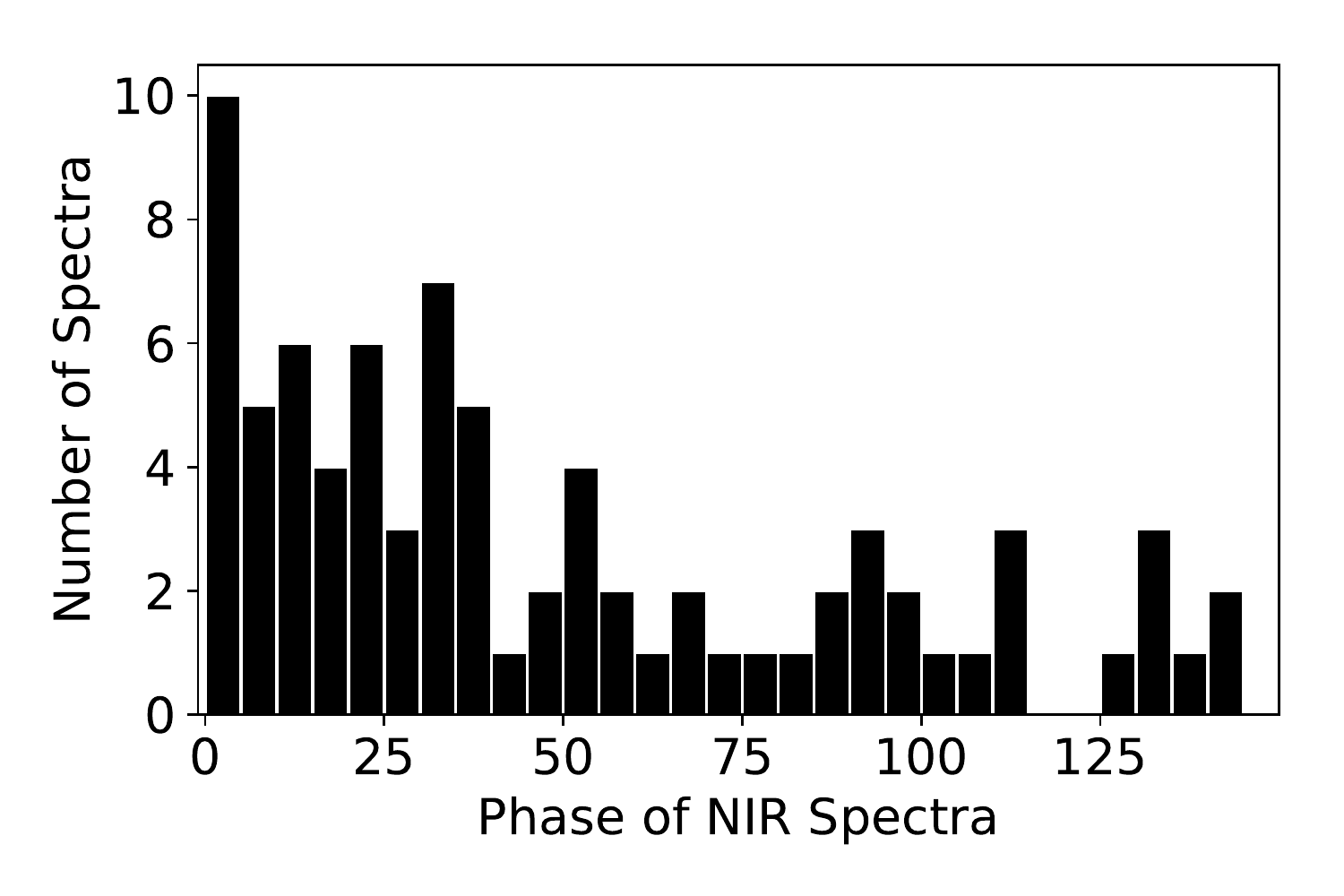}
    \caption{Number of SNe~II with time-series observations (top panel), and number of NIR spectra at each epoch relative to explosion date (bottom panel). We do not consider nebular phase spectra in this work.}
    \label{fig:firehistos}
\end{figure}

\begin{figure*}[b!]
    \centering
    \includegraphics[width=0.85\textwidth]{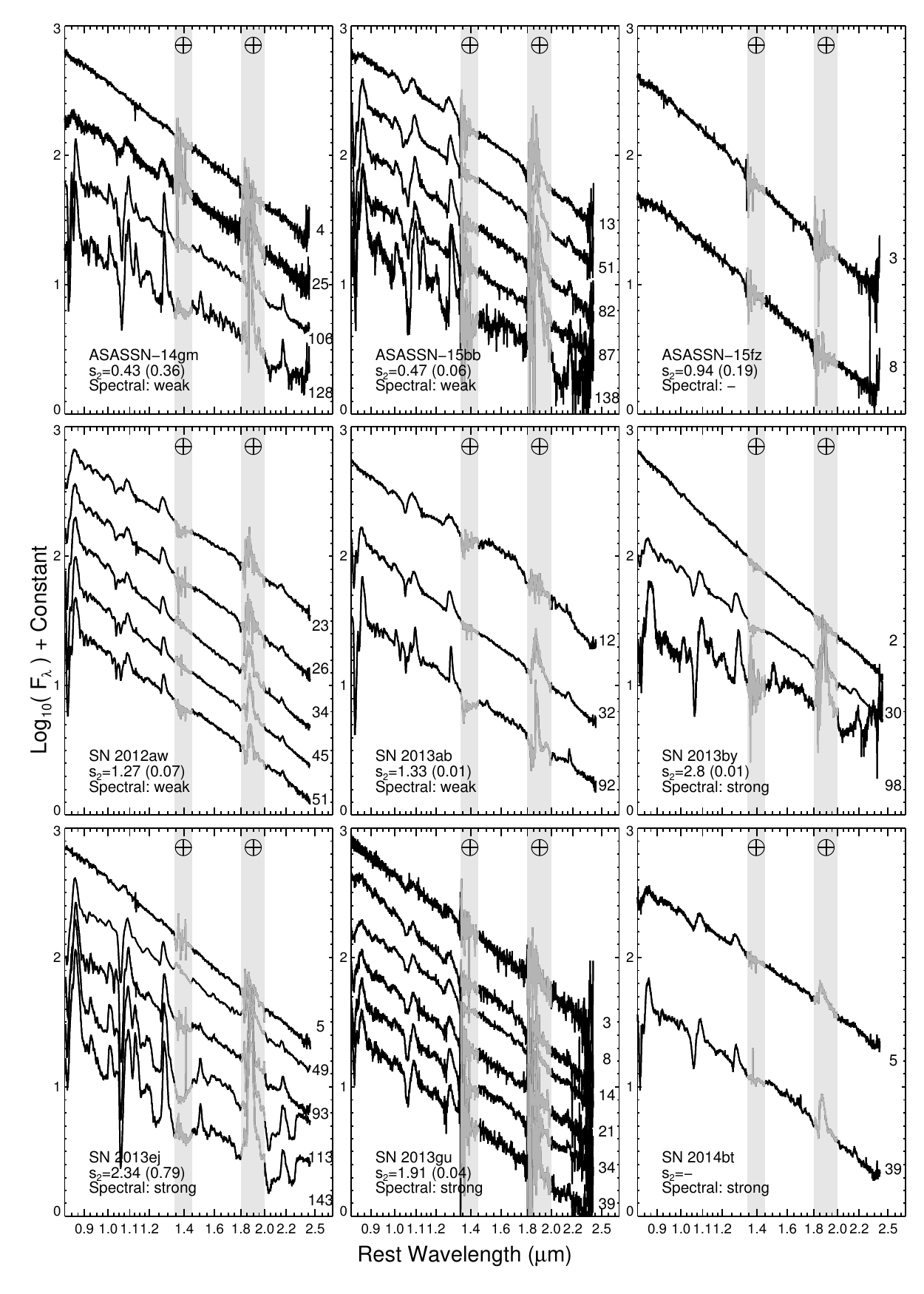}
    \caption{Time evolution for all SNe~II with two or more spectra from our sample, excluding SNe~2013gd and 2013hj which are displayed in Figure \ref{fig:13gd13hj}. Grey shaded areas mark wavelengths that have high telluric absorption from the atmosphere. The spectroscopic classification and photometric value $s_2$ is listed for each SN, with the $s_2$ error given in parentheses. Days since explosion is provided on the right side of the spectra.}
    \label{fig:nirPanels}
\end{figure*}

\begin{figure*}[b!]
    \centering
    \includegraphics[width=0.85\textwidth]{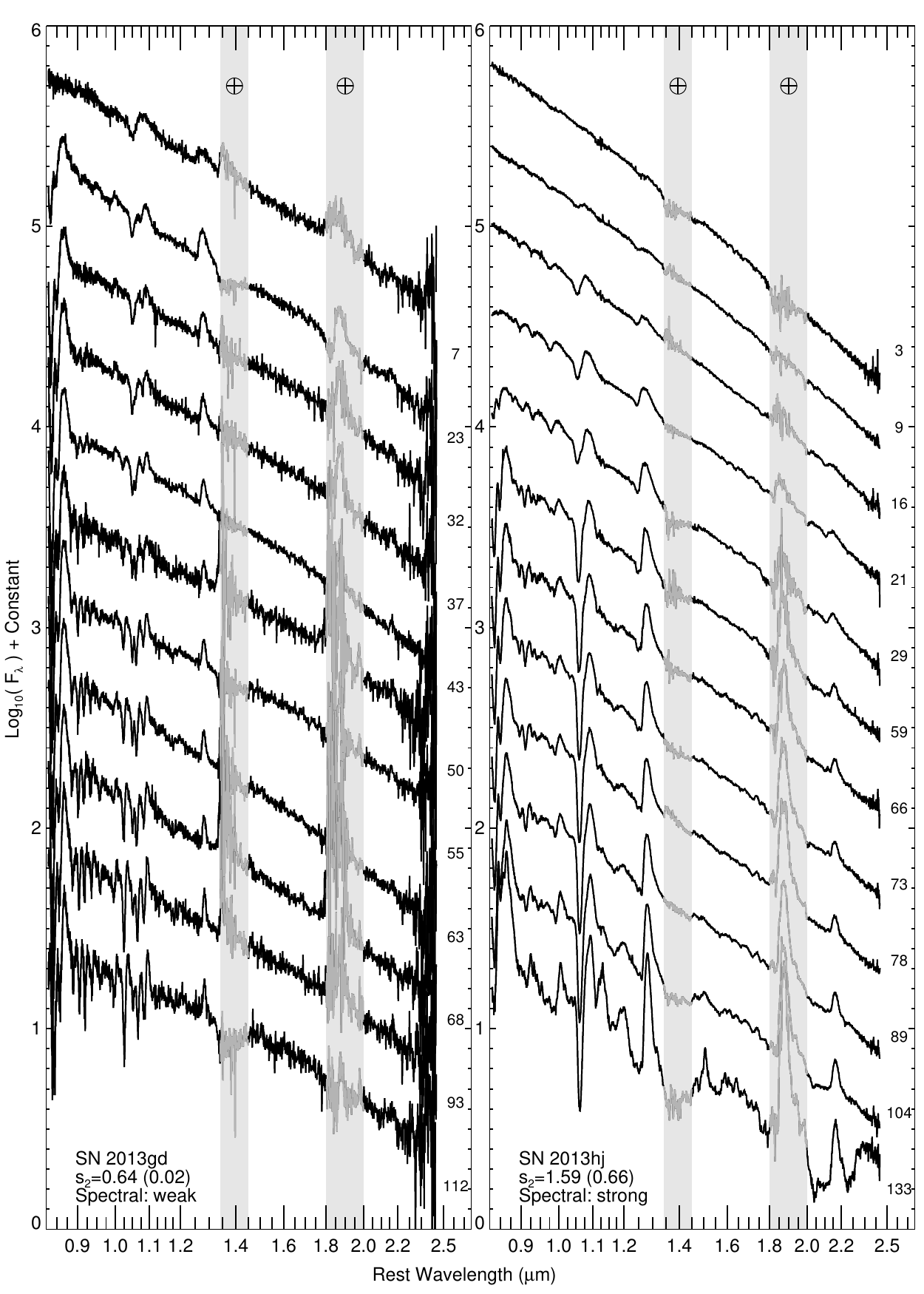}
    \caption{Same as Figure \ref{fig:nirPanels} but for SNe~2013hj and 2013gd, the two most complete time series in the sample.}
    \label{fig:13gd13hj}
\end{figure*}

\begin{figure*}
    \centering
    \includegraphics[width=0.9\textwidth]{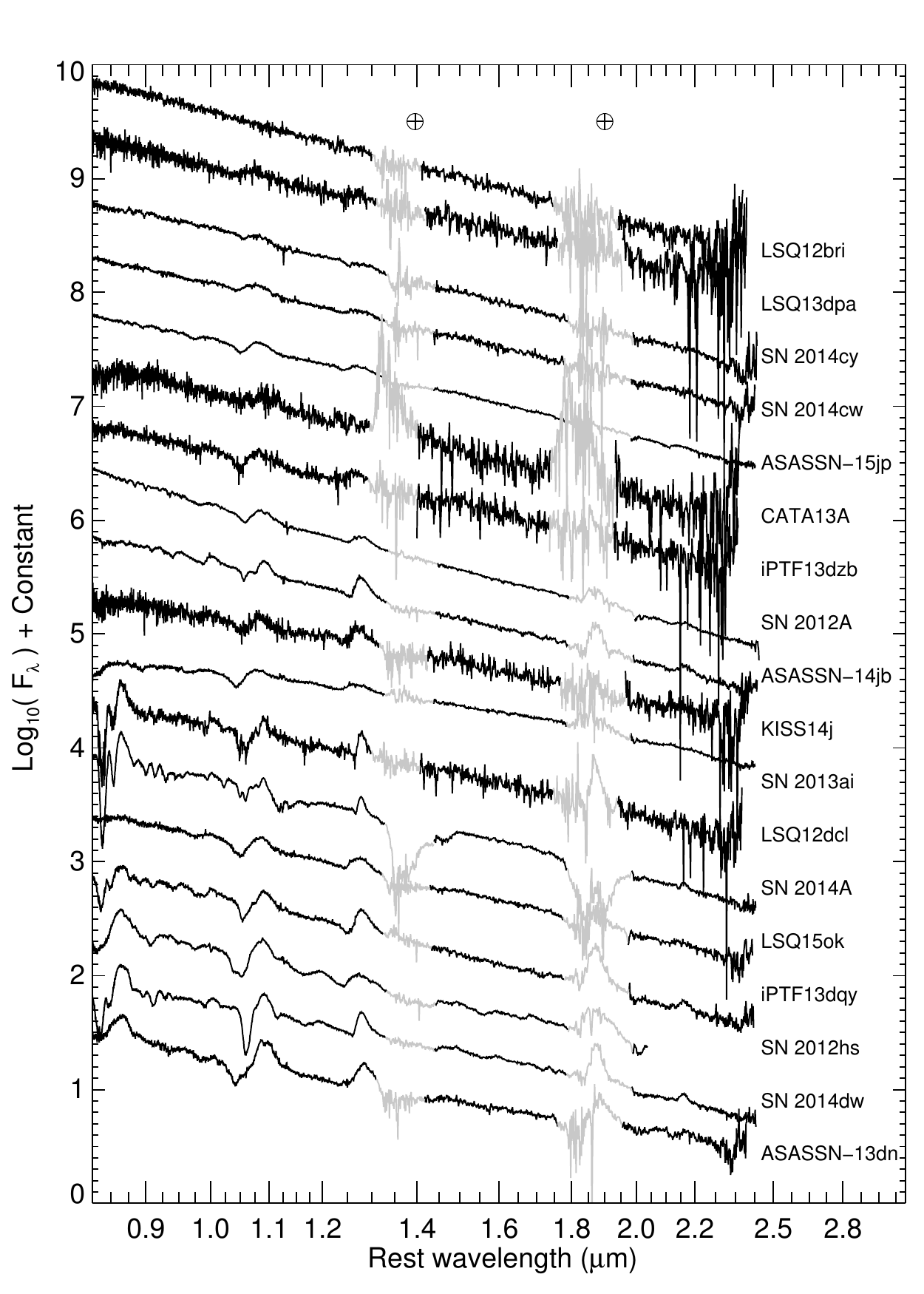}
    \caption{All SNe~II with only one spectrum from the CSP-II sample. Wavelengths that have high telluric absorption are plotted in grey. Explosion date for most SNe are uncertain, and thus, the spectra are grouped by similarity of features present.}
    \label{fig:nirSingle}
\end{figure*}

The CSP-II sample contains 92 NIR spectra of 32 SNe~II. Figure \ref{fig:firehistos} shows the phase of each NIR spectrum, as well as the distribution of number of spectra taken per SN. NIR spectra were obtained with the Magellan Baade Telescope, equipped with the Folded-port Infrared Echellette \citep[FIRE;][]{2013PASP..125..270S}, and with the NASA Infrared Telescope Facility (IRTF), equipped with SpeX \citep{2003PASP..115..362R}. The spectra were reduced and corrected for telluric absorption following the procedures outlined in \cite{2019PASP..131a4002H}. 
Spectra obtained after 300 days are not included in this sample because they are well into the nebular phase, giving a total of 81 NIR spectra from 30 SNe~II. See Table \ref{tab:sample} for a list of all SNe used, Table \ref{tab:NIR} for a log of the observations, and Figures \ref{fig:nirPanels}-\ref{fig:nirSingle} for all the spectroscopic data. Spectra can be downloaded from the Web. \footnote{https://csp.obs.carnegiescience.edu/data} The high throughput of FIRE allows us to recover enough counts in the telluric regions to enable telluric corrections, such as the P$\alpha$ feature, to a precision of 10\% or better in 70\% of our sample. Eleven SNe~II have multi-epoch observations, and 14 have well-sampled light curves. Photometry was obtained with the Las Campanas Observatory Swope and du Pont telescopes. The observing strategy and technique for the photometric data are described in \cite{2019PASP..131a4001P}.

The sample ranges from redshift 0.002 to 0.037, with a mean of 0.013 and median of 0.009. Table \ref{tab:hosts} lists the redshift and host galaxy for each SN~II within the sample. Host information obtained from the NASA/IPAC Extragalatic Database (NED)\footnote{http://ned.ipac.caltech.edu}, provides us with distance estimates to each SN~II. Table \ref{tab:distances} lists both redshift-independent and redshift-dependent distances. We use the redshift-independent distance to calculate absolute luminosity whenever possible. Our sample contains no SNe~IIn, SNe~IIb, or SN~1987A-like objects.

For each SN~II, the explosion date is assumed to be the midpoint between the last non-detection and discovery, while the uncertainty in the explosion date is taken as half of the range between last non-detection and the discovery date, a method adopted by \citet{2014ApJ...786...67A,2017ApJ...850...89G}. Spectral matching techniques are employed in order to constrain the explosion date using the Supernova Identification (SNID) code for each SN in the sample, even those with a well constrained, less than 20 days before discovery, last non-detection \citep{2007ApJ...666.1024B}. The phases obtained using SNID matches those from the last non-detection method, without any obvious bias toward last non-detection or discovery. It is assumed that there is a flat prior in the phase distribution. SNID does not have NIR templates, so all spectral matching was done in the optical with previously published data. Using SNID with updated SN~II templates has been shown to be a valid method to constrain the explosion epoch \cite[e.g.][]{2014ApJ...786...67A,2017ApJ...850...89G}. When using SNID, the explosion date is taken as the average of the best matched spectra and the range is taken as the error. Of the 30 SNe~II, 13 have well-defined last non-detections, 8 have explosion dates constrained with SNID, 4 have poorly constrained last non-detections, and 5 have no optical data and no available last non-detection date. 

\section{Measurements} \label{measurements}
\subsection{Photometric measurements} \label{photo}

Photometry is a valuable tool in characterizing SNe~II, allowing for comparisons between photometric and spectral properties. $V$-band light curves have become the standard in measuring photometric properties of SNe~II \citep[e.g.][]{2014ApJ...786...67A,2016MNRAS.459.3939V,2017ApJ...850...90G}. We follow the methods of \cite{2014ApJ...786...67A} and \cite{2014MNRAS.445..554F} for parameterizing the $V$-band light curves of SNe~II. For each light curve we attempt to measure $s_{\rm 1}$, $s_{\rm 2}$, $s_{\rm 3}$, $M_{\rm max}$, $M_{\rm end}$, $M_{\rm tail}$, $t_{\rm 0}$, $t_{\rm tran}$, $t_{\rm end}$, and $t_{\rm PT}$. Further discussion of the fitting techniques used can be found in Appendix \ref{appendix}. 

For our sample we use a cutoff of $s_{2}=1.4$ mag per 100 days, where $s_{2}$ is the slope of the SN light curve during the plateau phase, for the separation between slow and fast declining SNe~II. This number was chosen as it follows the division used in the literature to separate the IIP and IIL classes \citep{2014ApJ...787..139D,2015MNRAS.448.2608V,2016MNRAS.459.3939V,2016MNRAS.455.2712B}. It has been shown that there is no clear distinction between slow and fast declining SNe~II when looking at the slope of the plateau \citep{2014ApJ...786...67A,2017ApJ...850...90G}, however, a division has been made for this sample in order to compare spectral properties among the two photometric groups as a possible photometric separation has been seen by \cite{2012ApJ...756L..30A}.

\subsection{Spectroscopic Measurements} \label{spectro}

The expansion velocity and pseudo equivalent widths (pEW) of the NIR hydrogen features were measured for each SN~II for which photometric properties can be extracted. Both velocity and pEW measurements were performed by Gaussian fits to the absorption and emission features of each P Cygni profile. The region was manually selected for each line. The continuum in the selected region was estimated by a straight line. When fitting a single feature a straight line approximation of the continuum is often adequate. Localized blackbody fits to the same region were prone to over or under estimating the continuum due to contaminating features. Blackbody fits are also inaccurate past $\sim2\,\mu$m as free-free emission dominates the continuum. The region selected is then flattened by dividing the spectrum by the continuum. The velocity is measured by fitting for the minimum within the selected region. The pEW is measured via the method outlined in \cite{2004NewAR..48..623F,2007A&A...470..411G}.

Each spectrum used for velocity and pEW measurements has an error spectrum that represents our estimates of the flux measurement error at each pixel. For FIRE observations, the error spectra are the measured dispersion of the flux from mutiple frames at each pixel \citep{2019PASP..131a4002H}. For the rest of the observations, the error spectra are estimated by assuming that the Gaussian smoothed ($2\sigma$) spectra are the true spectral energy distribution (SED) of the SNe without errors. The flux errors are then taken as the standard deviation from the idealized SED.

A simple Gausssian function is used to fit a feature in order to determine the wavelengths of the feature minima and maxima. The resulting reduced $\chi^2$ is smaller than 1 in all fits. This indicates that a Gaussian function is not an accurate representation of the P-Cygni profile shape and perhaps the flux errors were underestimated. However, we consider this method adequate for the purpose of determining the feature extrema. A conservative velocity error is obtained by scaling the Gaussian fit error by the inverse of the reduced $\chi^2$. The diversity in the parent population is not considered in this error.

The pEW of each absorption and emission feature is directly measured by defining a straight-line continuum, and integrating over the enclosed continuum removed area, without assuming any functional form for the shape of the feature. The errors in the measurement of pEW were estimated via Monte Carlo, with realizations generated using the flux error spectra.

\section{Line Identifications} \label{lineID}

\begin{figure*}
    \centering
    \includegraphics[width=1.85\columnwidth]{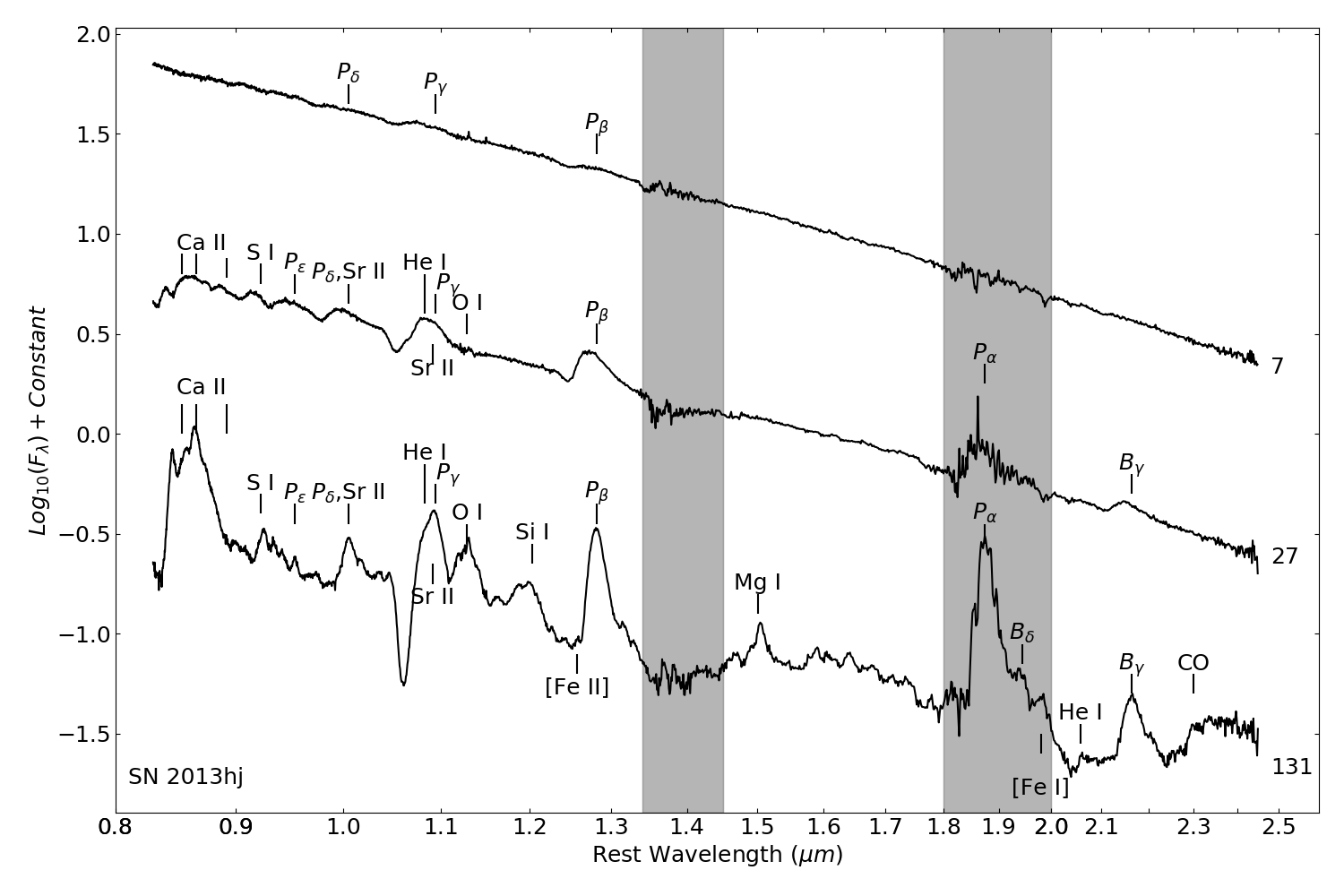}
    \caption{Time series spectra of SN~2013hj with the dominant ions labeled. The number to the right of each spectrum denotes the phase in days since explosion. The vertical lines mark the rest wavelength of each ion species. Regions of strong telluric absorption are outlined in grey.}
    \label{fig:labeled}
\end{figure*}

SNe~II as a whole are generally similar spectroscopically in the NIR. Using the spectral line identifications from \cite{1987ApJ...320L.121B} and \cite{1989MNRAS.238..193M}, we identify the features and evolution over time of the SNe~II within our sample; see Figure \ref{fig:labeled}.

\subsection{Early Phase}

Very early time spectra, less than $2-3$ days post explosion, exhibit few spectral lines, as can be seen in the top spectrum of Figure \ref{fig:labeled}, and can be well approximated by a blackbody Rayleigh-Jeans approximation in the NIR. At around 10 days past explosion, the hydrogen Paschen series begins to emerge with $P_{\gamma}$ $\lambda1.094\,\mu$m, and $P_{\beta}$ $\lambda1.282\,\mu$m the first features seen.

\subsection{Plateau Phase} \label{plateau}

As SNe~II settle into the plateau phase $\sim20$ days past explosion more features appear, see the middle spectrum in Figure \ref{fig:labeled}. The Ca\,{\footnotesize II} triplet is the strongest feature at this phase. It initially just shows an absorption feature and develops a stronger P Cygni profile as time passes. It is comprised of a blend of three Ca II lines: two close together and one more prominent, located at $\lambda\lambda0.854\,\mu$m, $0.866\,\mu$m, and $0.892\,\mu$m, respectively. The development of this feature is not uniform, with many SNe~II showing different line strengths.

The S\,{\footnotesize I} $\lambda0.922\,\mu$m feature shows up in the NIR around 30 days post explosion. The S\,{\footnotesize I} P Cygni profile strengthens over time in all SNe~II.

Sr\,{\footnotesize II} $\lambda1.004$ and $\lambda1.092\,\mu$m are both blended with H\,{\footnotesize I} features and cannot be resolved. Sr\,{\footnotesize II} $\lambda1.033\,\mu$m appears in all SNe~II but with a varied evolution with many SNe not showing Sr\,{\footnotesize II} until after the plateau phase ends; however, there are SNe which show strong Sr\,{\footnotesize II} absorption from Sr\,{\footnotesize II} $\lambda1.033\,\mu$m during the plateau. For these cases, the feature can be seen as early as $20$ days past explosion, increasing in strength over time and becoming a dominant feature in the SN spectrum.

$P_{\delta}$ $\lambda1.005\,\mu$m is one of the weaker hydrogen lines detectable in SNe~II NIR spectra. Blended with the $P_{\delta}$ feature is Sr\,{\footnotesize II} $\lambda1.004\,\mu$m that causes stronger line blending than is seen in the other hydrogen features. The strength of the P Cygni profile does not increase as quickly as other hydrogen features.

$P_{\gamma}$ $\lambda1.094\,\mu$m is highly blended with He\,{\footnotesize I} $\lambda1.083\, \mu$m and Sr\,{\footnotesize II} $\lambda1.092\, \mu$m. It has a very weak absorption, that is only noticeable in most SNe towards the end of the plateau as a notch on the red side of the He\,{\footnotesize I} absorption component. The emission in this region is attributed mostly to $P_{\gamma}$ and evolves similarly to the other H I features. 

\begin{figure}
    \centering
    \includegraphics[width=1.0\columnwidth]{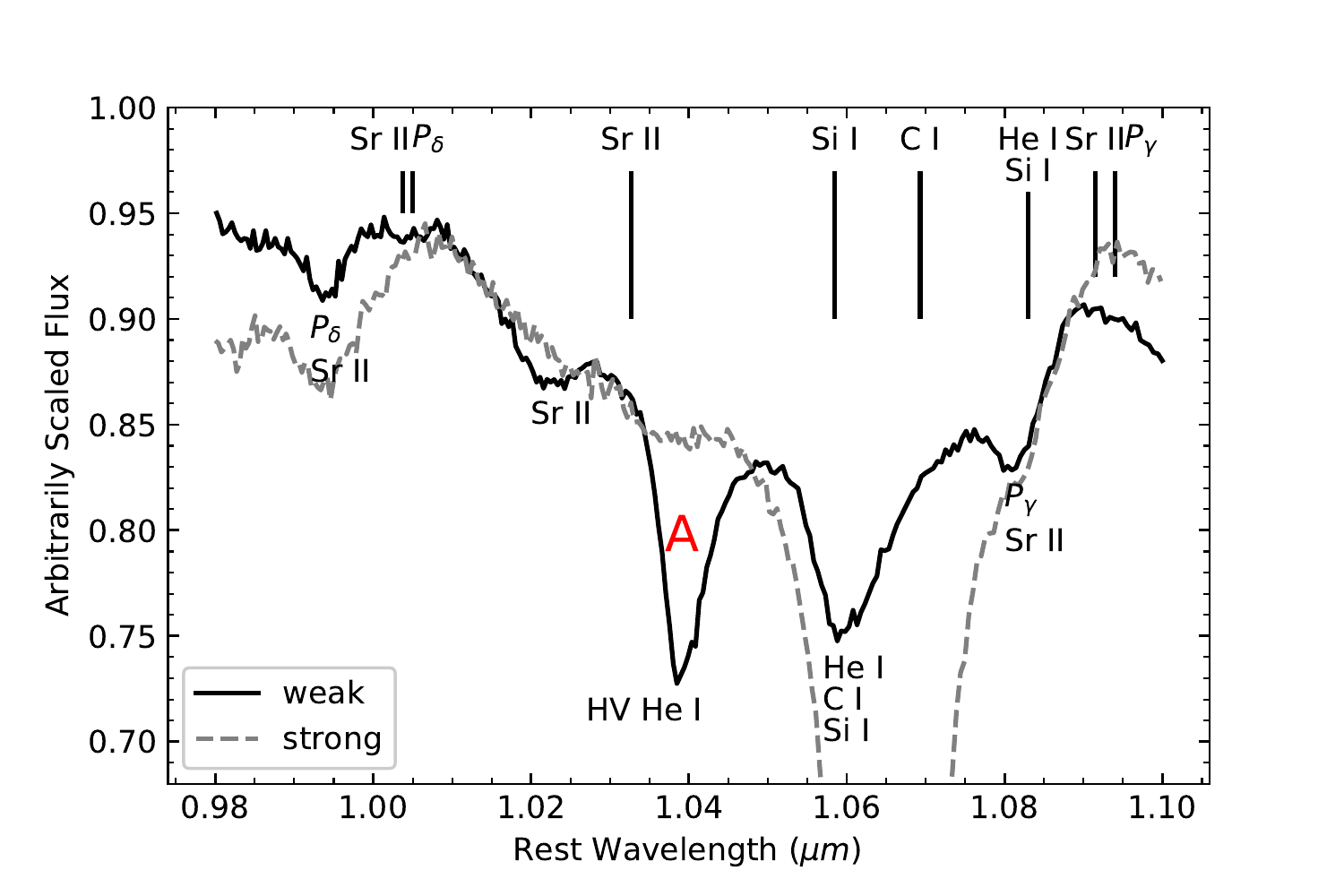}
    \caption{$\sim$30 day spectra of the \emph{weak} SN~2012aw (black) and the \emph{strong} SN~2013hj (grey) zoomed in on the region around He\,{\footnotesize I} $\lambda1.083\,\mu$m. The absorption and emission components of each line in the SN~2012aw spectrum are labeled with possible identifications. Emission features are marked at rest wavelength above the spectra. SN~2013hj is shown for comparison of line strengths and features present in the region. SN~2012aw and SN~2013hj are good representations of the weak and strong spectral classes, respectively. These features are discussed in detail in Section \ref{dichotomy}.}
    \label{fig:cartoon}
\end{figure}

During the plateau the most prominent absorption arises from He\,{\footnotesize I} $\lambda1.083\,\mu$m and has other possible contributions from Sr\,{\footnotesize II} $\lambda1.092\,\mu$m and $P_{\gamma}$ $\lambda1.095\,\mu$m. Since other Sr\,{\footnotesize II} transitions in this region are weak, it is unlikely that this absorption is dominated by Sr\,{\footnotesize II}. He\,{\footnotesize I} is excited non-thermally and would only be seen if the $^{56}Ni$ was located close to the He region in the ejecta \citep{1988ApJ...335L..53G}.
We note that this feature is not expected to have a significant contribution from H\,{\footnotesize I} because no strong H\,{\footnotesize I} absorption is present in the other Paschen series lines. 
However, the emission of this P Cygni profile is most likely to have contribution from $P_{\gamma}$, as strong emission features are seen in other Paschen series lines.

A strong He\,{\footnotesize I} absorption profile with a weak emission component is present in the early optical spectra of SNe~II \citep{2017ApJ...850...89G}. 
Thus in the NIR, we do not expect He\,{\footnotesize I} to contribute a large amount of flux in emission to this feature. 
It is also possible for the emission component to be shifted blueward by the absorption of H\,{\footnotesize I} and Sr\,{\footnotesize II}.
At 50 days past explosion, this P Cygni profile consists of a blend comprised of He\,{\footnotesize I}, H\,{\footnotesize I}, and Sr\,{\footnotesize II}, with He\,{\footnotesize I} $\lambda1.083\,\mu$m being the dominant contributor in absorption. Thus we will refer to this absorption as He\,{\footnotesize I} for the remainder of this work.
At later times, as the photosphere recedes to the inner hydrogen-rich region, we expect heavier elements to appear. Hence, late contributions from C\,{\footnotesize I} and Si\,{\footnotesize I} are likely.

Around 50 days post explosion a dichotomy in the region around He\,{\footnotesize I} $\lambda1.083\,\mu$m appears. Figure \ref{fig:cartoon} shows two SNe which represent the dichotomy. SNe with more prominent He\,{\footnotesize I} absorption show no other lines in this region until later phases when a small absorption, comprised of $P_{\gamma}$ and Sr\,{\footnotesize II} $\lambda1.092\,\mu$m appears on the red side of the He\,{\footnotesize I} absorption. SNe with a shallower He\,{\footnotesize I} absorption show other features which are not seen in SNe that exhibit a deeper He\,{\footnotesize I}. The $P_{\gamma}$/Sr\,{\footnotesize II} absorption appears earlier in these SNe and an additional absorption appears on the blue side of He\,{\footnotesize I}, hereafter feature A (Figure \ref{fig:cartoon}).

Feature A, if present, appears before the plateau and is similar to the He\,{\footnotesize I} $\lambda1.083\,\mu$m transition at these early times. During the plateau, feature A does not significantly increase in strength compared to the other absorption features in the spectrum and becomes less prominent compared to He\,{\footnotesize I} absorption in the same region.

O\,{\footnotesize I} $\lambda1.129\,\mu$m  appears on the red side of the $P_{\gamma}$ emission towards the end of the plateau phase. Si\,{\footnotesize I} $\lambda1.203\,\mu$m is also seen during this period and strengthens over time.

$P_{\beta}$ $\lambda1.282\,\mu$m is seen at the beginning of the plateau and  exhibits a symmetric P Cygni profile, that becomes dominated by its emission feature with very little absorption. Towards the end of the plateau, the $P_{\beta}$ line profile seems to widen, likely due to the presence of a Si I multiplet in the region.

$P_{\alpha}$ $\lambda1.875\,\mu$m is seen early during the plateau, appearing predominately in emission. It may form earlier, along with the other Paschen series lines; however, it is located in a band of strong telluric absorption making it difficult to study for most spectra. Looking at SNe~II with high signal-to-noise (S/N) ratios in this region, we see that $P_{\alpha}$ increases in strength over time. In the same band of telluric absorption, $P_{\epsilon}$ $\lambda0.954\,\mu$m and the Brackett series hydrogen line $B_{\delta}$ $\lambda1.944\,\mu$m appear around 30 days post explosion as weak emission features. These lines originate from the same upper level and thus both grow in strength with neither getting particularly strong compared to the other hydrogen emission features seen.

$B_{\gamma}$ $\lambda2.165\,\mu$m shows up around the same time as $P_{\alpha}$ and cannot be resolved if the signal-to-noise ratio (S/N) of the spectrum is low, e.g., in some of the SN~2013gd spectra. The absorption of $B_{\gamma}$ is weak enough that it can only be seen in the highest S/N spectra within the sample. The emission is also weak when compared to the Paschen-series.

\subsection{Radioactive Decay Tail}

After the plateau phase ends, we observe more metal lines forming, mostly in emission, and the emission component of all lines which formed during the plateau strengthen. 

[Fe\,{\footnotesize II]} $\lambda1.279\,\mu$m, Mg\,{\footnotesize I} $\lambda1.503\,\mu$m and [Fe\,{\footnotesize I]} $\lambda1.980\,\mu$m emerge during the radioactive decay tail as weak emission features. Some SNe, e.g. SN~2013gd, do not show significant Mg\,{\footnotesize I} emission at any time. He\,{\footnotesize I} $\lambda2.058\,\mu$m appears as a weak absorption and does not appear in all SNe~II. This He\,{\footnotesize I} $\lambda2.058\,\mu$m transition should be highly correlated with the He\,{\footnotesize I} $\lambda1.083\,\mu$m transition.

CO begins to appear after the plateau. The first CO overtone appears largely as emission features around $2.3\, \mu$m. The earliest detection in this data set is found in the SN~2013by spectrum taken 95 days post explosion. The CO feature becomes stronger over time, appearing in more SNe~II around 120 days post explosion. We find CO in four SNe~II within our sample, appearing in most spectrum taken after 120 days post explosion. 

\begin{figure*}
    \centering
    \includegraphics[width=1.75\columnwidth]{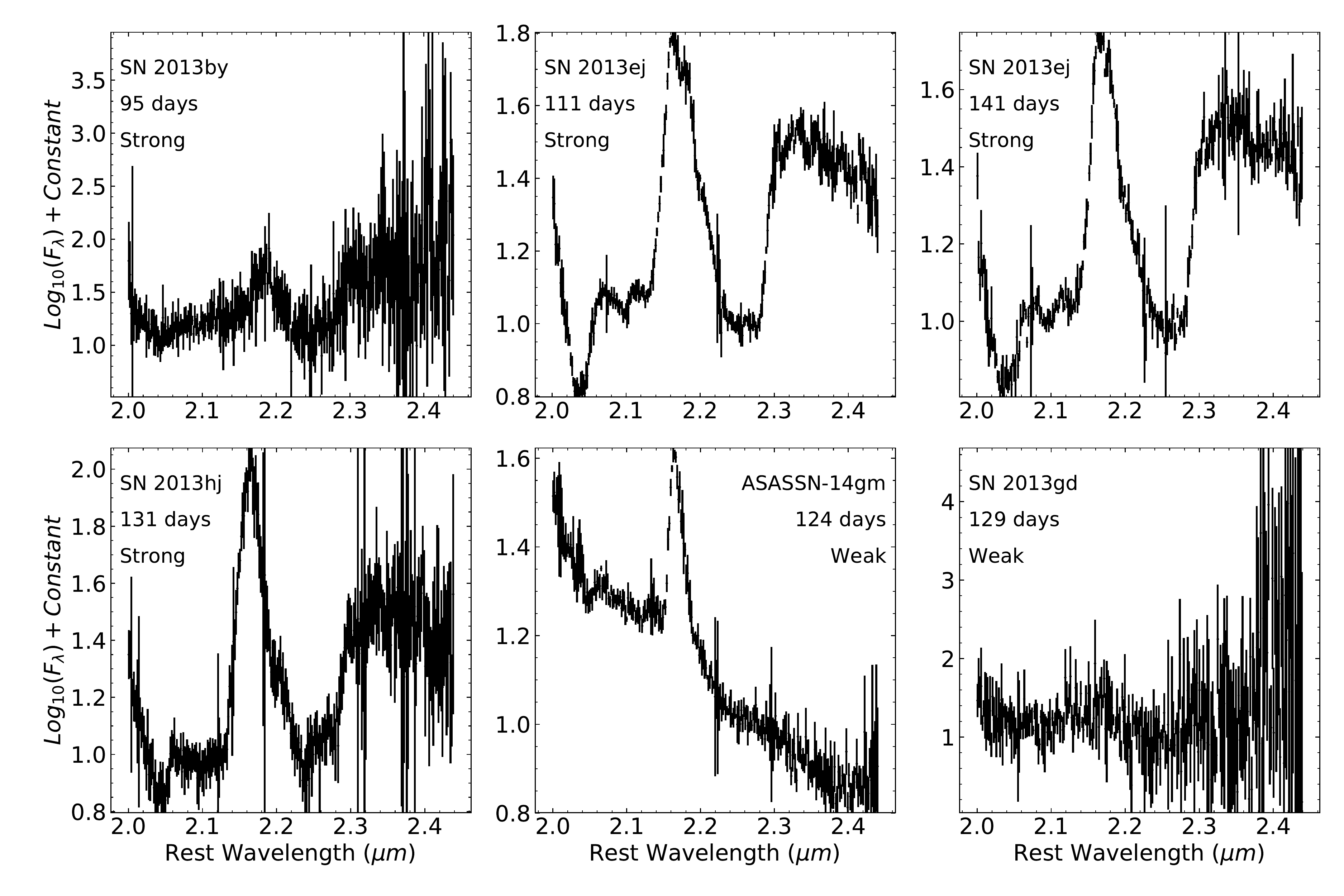}
    \caption{Late time spectra zoomed in on the region where CO is expected, around $2.35\,\mu$m. ASASSN14-gm and SN~2013gd have the weakest CO features (if present at all) past 120 days post explosion seen in the CSP-II sample.}
    \label{fig:CO}
\end{figure*}

Late time spectra with possible CO detections are shown in Figure \ref{fig:CO}. Modeling and further analysis of CO will be performed in future work.

\section{Feature Measurements} \label{quantMeasure}

The NIR spectral features were measured as outlined in Section \ref{spectro} and are presented below. Figures \ref{fig:velEvo} and \ref{fig:pewEvo} show the velocity and pEW evolution for the CSP-II sample over time, respectively. Figures \ref{fig:velEvo} and \ref{fig:pewEvo} present the data split by $s_{2}$ in order to search for any relationships between photometric and NIR spectral properties. 

\begin{figure*}
    \includegraphics[width=\textwidth]{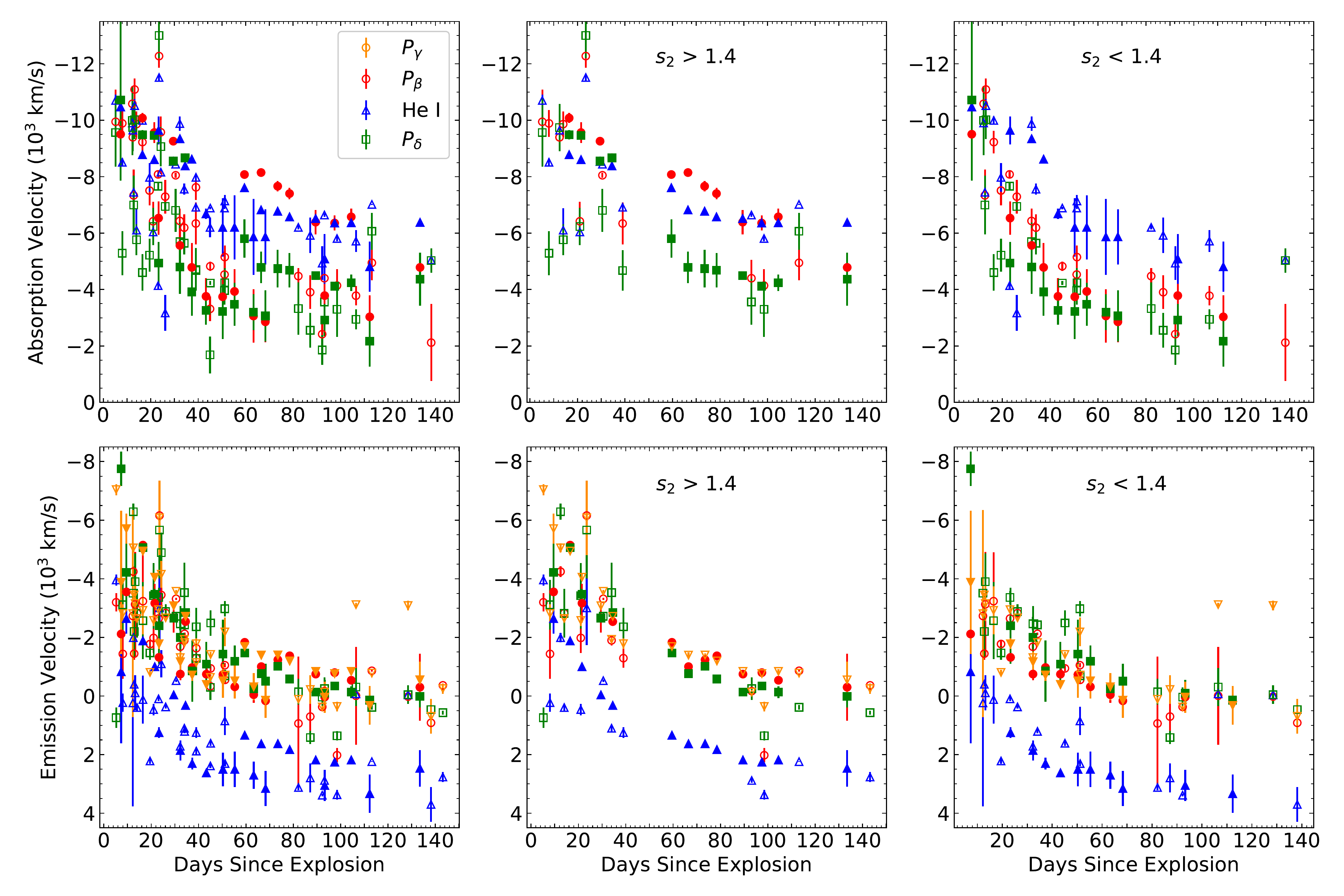}
    \caption{Evolution of the absorption and emission NIR velocitiy features of SNe~II. The first column shows all of the CSP-II data together. The last two columns show the data split by $s_2$ value as defined in Section \ref{photo}. Velocity measurements of He\,{\footnotesize I} $\lambda1.083\,\mu$m are also included. The velocity evolution of the emission features is similar to what was seen by \citet{2014MNRAS.441..671A}. The fast declining SN~2013hj and slow declining SN~2013gd are represented by filled points.}
    \label{fig:velEvo}
\end{figure*}

\begin{figure*}
    \centering
    \includegraphics[width=1.75\columnwidth]{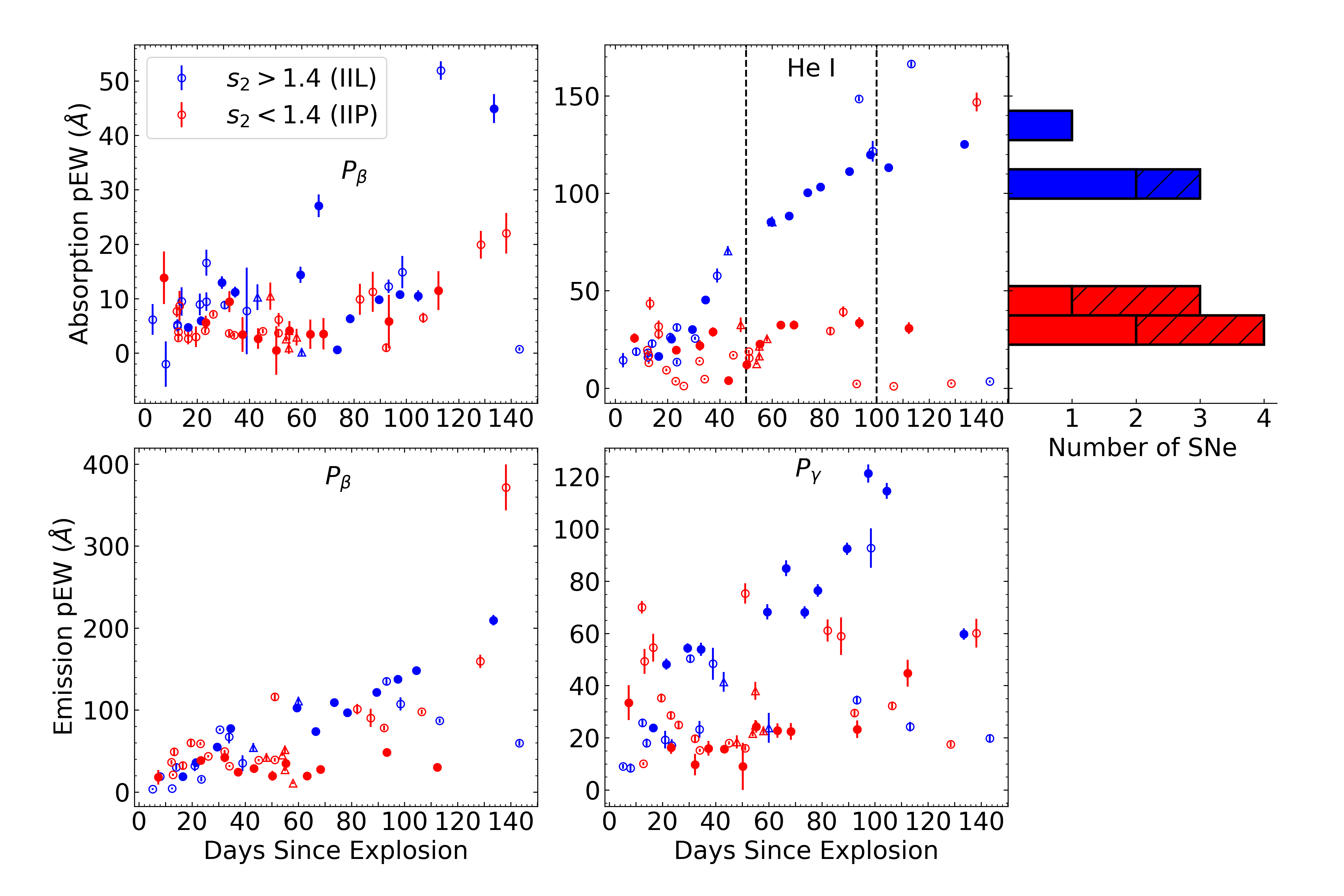}
    \caption{Evolution of the absorption and emission pEW width of SNe~II NIR features over time. Included are pEW measurements of He\,{\footnotesize I} $\lambda1.083\,\mu$m. There is a stark contrast between slow and fast declining SNe~II with regards to He\,{\footnotesize I} absorption pEW. The emission pEW also splits with ASASSN-15bb being the sole slow declining SN~II lying with the fast declining SNe~II curve. The fast declining SN~2013hj and slow declining SN~2013gd are represented by filled circles. SNe not from CSP-II are plotted as open triangles. The vertical black lines drawn in the upper right panel represent the time in which a SN~II with $s_{\rm 2}=1.4$ should be in its plateau phase, when a NIR classification can be made. The histograms show how many SNe lie between 50 and 100 days past explosion. SNe not from CSP-II are plotted with hatched patterns in the histograms. The histogram shows all data between $50-100$ days interpolated to $75$ days.}
    \label{fig:pewEvo}
\end{figure*}

He\,{\footnotesize I} $\lambda1.083\,\mu$m strengthens over time until nebular phases and often has the highest absorption pEW, increasing from $10$ \AA{} to $\sim50-100$ \AA{}, in the NIR spectrum. There is a dichotomy seen in the He\,{\footnotesize I} pEW, SNe well above $50$ \AA{} and those below $50$ (Figure \ref{fig:pewEvo} top right panel). The velocity of He\,{\footnotesize I} absorption decreases over time, similar to the other spectral features, going from $\sim11000$ km s\textsuperscript{-1} at early times to $\sim 4000$ km s\textsuperscript{-1} after the plateau phase ends.

The velocity of the $P_{\beta}$ absorption at early and late times is nearly uniform among the sample, however, during the plateau there is a split in the velocities between fast and slow decliners, $\sim$7500 km s\textsuperscript{-1} for fast decliners and $\sim$4500 km s\textsuperscript{-1} for slow decliners. The absorption pEW increases from $\sim1-2$ \AA{} to $10$ \AA{} during the plateau. The emission pEW increases to over $100$ \AA{} during the plateau, becoming the second strongest emission, behind $P_{\alpha}$, during this period. The emission velocities and absorption/emission pEWs show no correlation with photometric subclass.

The absorption and emission equivalent widths of the $P_{\delta}$ P Cygni profile do not increase as quickly as other hydrogen features. With velocities around $11000$ km s\textsuperscript{-1}, $P_{\delta}$ often has the lowest velocities of all present hydrogen features, especially in fast declining SNe where the $P_{\beta}$ velocities are higher, at certain epochs, than in slow declining SNe. The $P_{\delta}$ velocities are likely influenced by the Sr\,{\footnotesize II} $\lambda1.092\,\mu$m blend causing them to be lower than expected. The $P_{\delta}$ absorption pEW evolves much like $P_{\beta}$ going from $\sim1-2$ \AA{} to $10$ \AA{} during the plateau. The emission pEW is uniform amongst the sample, evolving linearly over time from $\sim1-2$ \AA{} to $50$ \AA{} at $120$ days past explosion.

\begin{figure}
    \centering
    \includegraphics[width=\columnwidth]{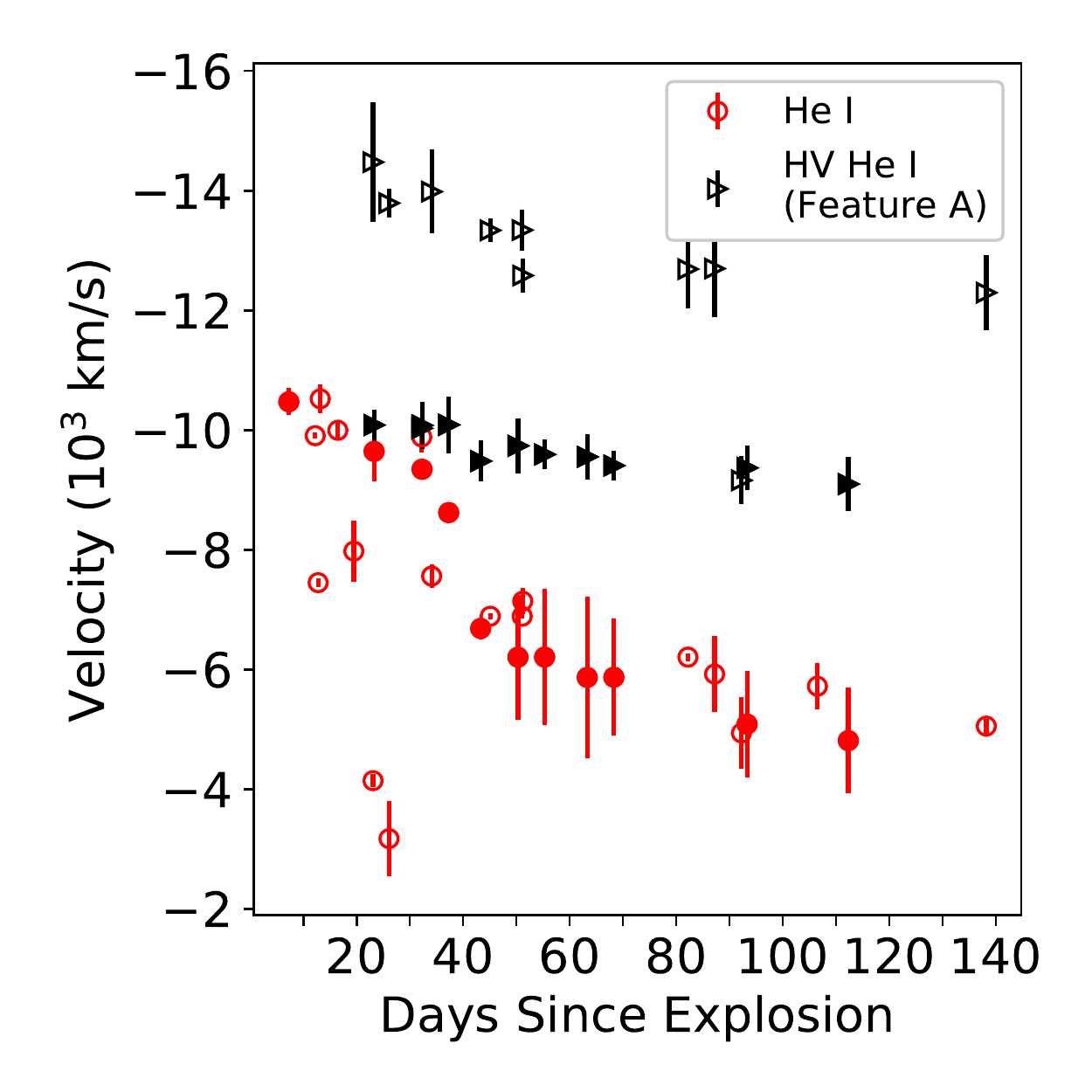}
    \caption{Velocity comparison of feature A and the He\,{\footnotesize I} $\lambda1.083\,\mu$m absorption for each SN in which both profiles are present. These velocities assume the rest wavelength of He\,{\footnotesize I} $\lambda1.083\,\mu$m. SN~2013gd is represented by filled points. Feature A exhibits a flat velocity evolution, unlike any other feature in a SN~II spectrum. Feature A also has a large velocity spread, as low as $10000$ km s\textsuperscript{-1} and as high as $14000$ km s\textsuperscript{-1}. Feature A is most likely HV He\,{\footnotesize I}.}
    \label{fig:FeatureA}
\end{figure}

Feature A has the most inconsistent evolution, in velocity and pEW, compared to other NIR features (Figure \ref{fig:FeatureA}). This feature has been previously identified as either Si\,{\footnotesize I} $\lambda1.033\,\mu$m or high-velocity He\,{\footnotesize I} $\lambda1.083\,\mu$m \citep{2014MNRAS.438..368T}. The absence of other metal lines in the spectra at similar times suggests that feature A is not Si\,{\footnotesize I}. Furthermore, the timing of the feature also suggests it is not Si as the photosphere is not expected to be in the metal rich region or the inner region of the hydrogen-rich envelope at these early epochs. Therefore, we conclude that feature A is most likely high-velocity He\,{\footnotesize I}. If present in a SN, it appears early, for example 7 days past explosion in SN~2013gd, and does not change significantly over time in pEW or velocity (Figure \ref{fig:classificationSpec}). 
The early onset of the feature suggests that it was produced in the outermost layer of the envelope.
Unlike the photospheric He\,{\footnotesize I} absorption, feature A does not appear at the same wavelength in each SN, if it appears at all. It has a velocity spread between objects of $\sim 4000$ km s\textsuperscript{-1} at similar epochs. This velocity spread assumes the feature is He\,{\footnotesize I}.

\section{Observed NIR Dichotomy} \label{dichotomy}

NIR spectra of SNe~II are generally uniform showing the same features with similar time evolution between SNe. However, they exhibit a dichotomy of properties that emerges around 50 days post explosion and lasts until the end of the SN plateau phase. This is most prominent in the region around He\,{\footnotesize I} $\lambda1.083\,\mu$m; a comparison between the two groups is shown schematically in Figure \ref{fig:cartoon}. Therefore, we classified our sample of SNe~II based on this dichotomy and describe the differences between the two groups (\emph{weak} and \emph{strong}) below.

\begin{figure*}
    \centering
    \includegraphics[width=1.75\columnwidth]{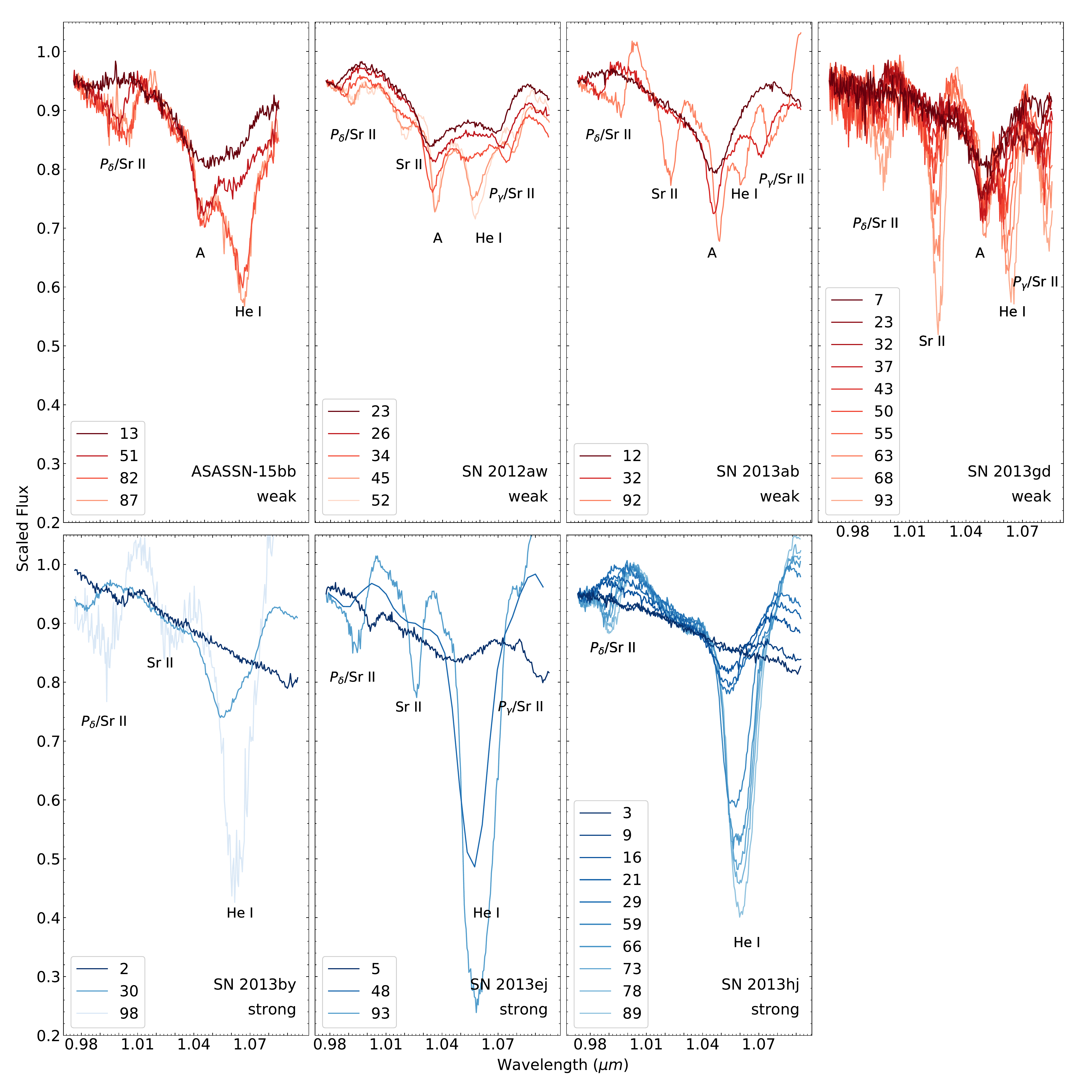}
    \caption{All SNe from the CSP-II sample that could be classified as either \emph{strong} or \emph{weak}. \emph{Weak} SNe are plotted in red and \emph{strong} SNe are plotted in blue. Ion names mark the most likely dominant species for each absorption feature. The \emph{weak} SNe exhibit a range of strength and velocity evolutions within the subclass, however, can still be defined by weak He\,{\footnotesize I} absorption and the presence of feature A. Feature A is most likely HV He\,{\footnotesize I}.}
    \label{fig:classificationSpec}
\end{figure*}

We define \emph{weak} SNe~II as those with He\,{\footnotesize I} $\lambda1.083\,\mu$m absorption pEW less than $50$ \AA{} after $50$ days from explosion and \emph{strong} SNe~II with He\,{\footnotesize I} pEW greater than $50$ \AA{}. This quantitative division was arbitrarily chosen based on Figure \ref{fig:pewEvo}. There are no SNe with intermediate pEW ($50-75$ \AA{}) in our sample (histogram in the top right panel of Figure \ref{fig:pewEvo} shows the pEW of each SN interpolated to 75 days). Figure \ref{fig:classificationSpec} shows all the SNe in the sample that could be classified as \emph{weak} or \emph{strong}. Although the pEW of He\,{\footnotesize I} was used to classify the two groups, several other differences between the two groups are also observed.

\emph{Weak} SNe show an accompanying absorption feature, feature A, on the blue side of He\,{\footnotesize I}, and \emph{strong} SNe do not. As discussed in Section \ref{plateau}, feature A is most likely high velocity He\,{\footnotesize I} $1.083\,\mu$m. Feature A always shows up before photospheric He\,{\footnotesize I} $1.083\,\mu$m, consistent with the interpretation that feature A is HV He\,{\footnotesize I}. \emph{Weak} SNe show earlier notches arising from $P_{\gamma}$/Sr\,{\footnotesize II} absorption on the blue side of the He\,{\footnotesize I}/H\,{\footnotesize I} emission, $\sim$20 days past explosion. \emph{Strong} SNe show the $P_{\gamma}$/Sr\,{\footnotesize II} absorption at $\sim40$ days. \emph{Weak} SNe tend to exhibit strong Sr II features in the $1.0-1.1\,\mu$m region sooner, $\sim20$ days. \emph{Strong} SNe may not show Sr II at all, e.g. SN~2013hj. \emph{Weak} SNe have a more layered velocity structure and decline in velocity more quickly than \emph{strong} SNe (Figure \ref{fig:velEvo}). CO emission can be seen in \emph{strong} SNe before 100 days past explosion. Emission from the first CO overtone appears at later times, past 120 days, in \emph{weak} SNe, if seen at all.

\begin{figure*}
    \centering
    \includegraphics[width=1.5\columnwidth]{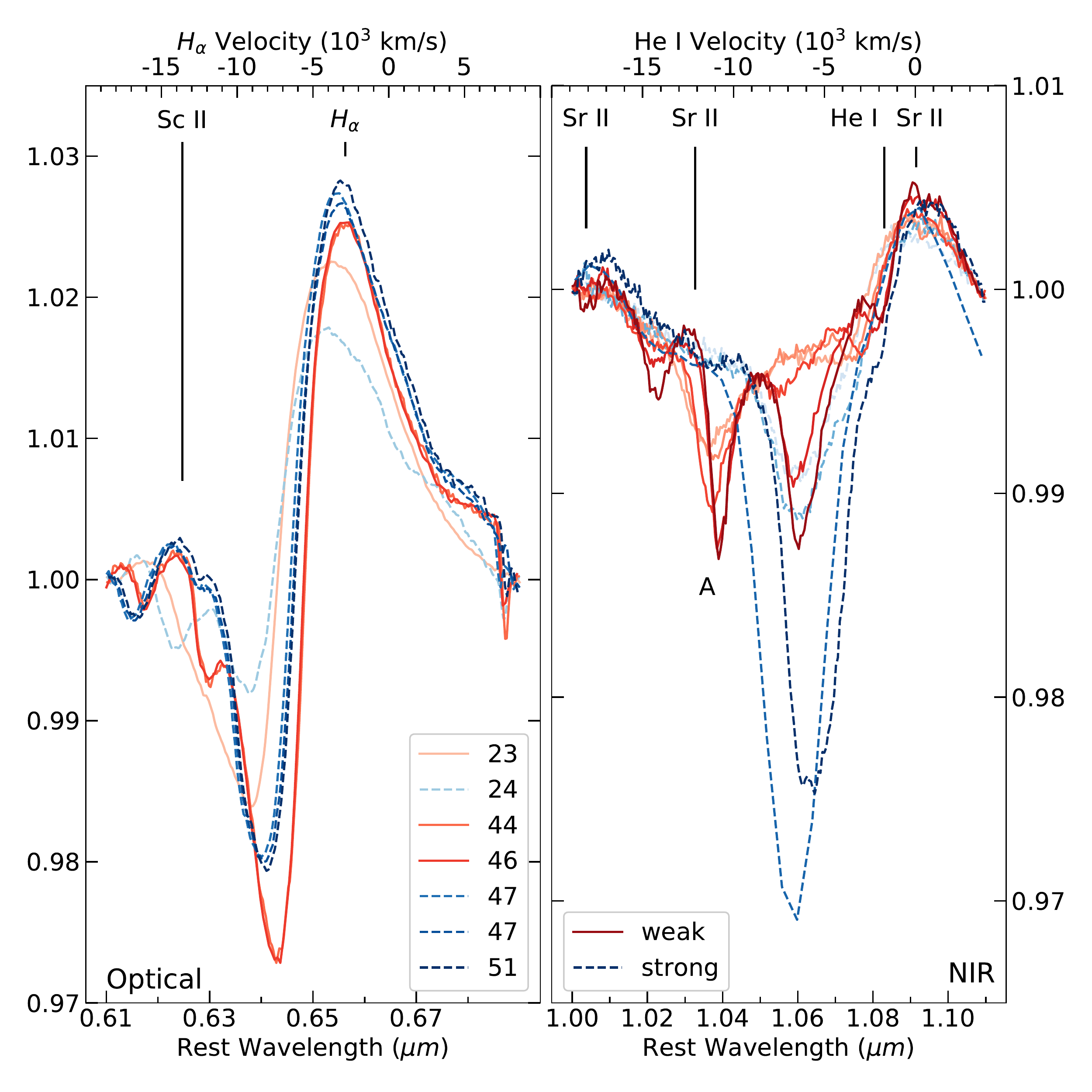}
    \caption{Comparison of the optical (left) and NIR (right) data of the two spectral classes. The \emph{weak} spectral class is plotted in red with darker shades representing later phases. The \emph{strong} spectral class is plotted in blue with darker shades representing later phases. The phase of each spectrum, rounded to the nearest day, is listed in the left panel. Both spectral types are plotted at similar phases. In the optical SN~2013ej (\emph{strong}) and SN~2012aw (\emph{weak}) are plotted. In the NIR SN~2013hj (\emph{strong}) and SN~2012aw (\emph{weak}) are plotted. SN~2013hj has little optical data but is spectroscopically similar to SN~2013ej in the NIR so we use SN~2013ej as the \emph{strong} spectral class representation in the optical. The difference between these two classes is much more obvious in the NIR.}
    \label{fig:nirOptCompare}
\end{figure*}

In the optical, a small notch on the blue of the $H_{\alpha}$ absorption was interpreted as HV H\,{\footnotesize I}, after 30 days post explosion, and its evolution is well-studied \citep{2017ApJ...850...89G}.
Following \cite{2014MNRAS.438..368T}, we compare the evolution of these HV features in Figure \ref{fig:nirOptCompare} to determine if the features are formed in the same region. 
In our sample only one SN~II shows both HV features: SN~2012aw.
However, this does not mean that the HV H\,{\footnotesize I} is absent in all other SNe~II, as it could be mixed into $H_{\alpha}$ as suggested by \cite{2007ApJ...662.1136C}. H\,{\footnotesize I} is also easier to excite than He\,{\footnotesize I}, which could explain the higher incidence of HV H\,{\footnotesize I} in the optical. It is possible for a SN to show a feature which looks similar to HV H\,{\footnotesize I} and no NIR HV counterpart, e.g. SN~2013ej. The absorption on the blue side of $H_{\alpha}$ in the optical may also be due to Si\,{\footnotesize II} as suggested by \cite{2014MNRAS.438L.101V} for SN~2013ej.
When these HV components appear in the same SNe, their velocities match, e.g. SN~2012aw with HV H\,{\footnotesize I} and HV He\,{\footnotesize I} both around $14000$ km s\textsuperscript{-1}.

The NIR spectroscopic classification of \emph{weak} and \emph{strong} are found to have a one-to-one correspondence with the IIP and IIL subclasses, based on the decline rate $s_{2}$.
For the SNe in our sample where we can determine a photometric classification (5 slow decliners and 4 fast decliners), all slow decliners are \emph{weak} and all fast decliners are \emph{strong}. 
When we include all of the previously published data that can be spectroscopically classified, either by pEW measurements or by the line profiles in the region around He\,{\footnotesize I} $\lambda1.083\,\mu$m (8 slow decliners and 2 fast decliners), we find only one exception to the rule, SN~2012A \citep{2013MNRAS.434.1636T}.
Looking at NIR spectra, the strong dichotomy suggests that fast declining and slow declining SNe~II are two distinct groups of objects, at least phenomenologically; whereas when viewed in the optical, these objects appear to have a continuous range of photometric and spectroscopic properties \citep{2014ApJ...786...67A,2017ApJ...850...89G,2017ApJ...850...90G,2019MNRAS.tmp.1836P}. A Silverman model \citep{1981JRSSB..43...97S} was used to create $s_{2}$ probability density functions (PDF) for both the CSP-II and \cite{2017ApJ...850...90G} photometric samples. A Silverman model assumes that each point can be represented by a Gaussian profile. These profiles can then be added together and normalized to create a PDF of the dataset. A PDF was made using the error from each $s_{2}$ measurement, however, for the majority of measurements this error is small compared to the $s_{2}$ value and produces a PDF that over-estimates the number of measurement modes. Thus, each point in the Silverman model is assigned the same kernel density that was chosen using the critical width function of \citet{1981JRSSB..43...97S}. The resulting PDFs of the two samples are similar despite the small numbers of the CSP-II sample, suggesting that the CSP-II sample is not photometrically biased.

\section{Principle Component Analysis} \label{PCA}

To further test the validity of the dichotomy observed in the NIR spectra of SNe~II, we performed a principal component analysis (PCA) on a selection of data within our sample. PCA has been applied in a variety of astronomical research previously \citep[e.g.][]{2006ApJS..163..110S,2007ApJ...663.1187H} to recognize patterns in the data and as a tool for creating spectral templates. PCA reduces the dimensionality of a multi-dimensional data set, such that a large fraction of the variance in the data is captured in a few principal components.

\begin{figure*}
    \includegraphics[width=\textwidth]{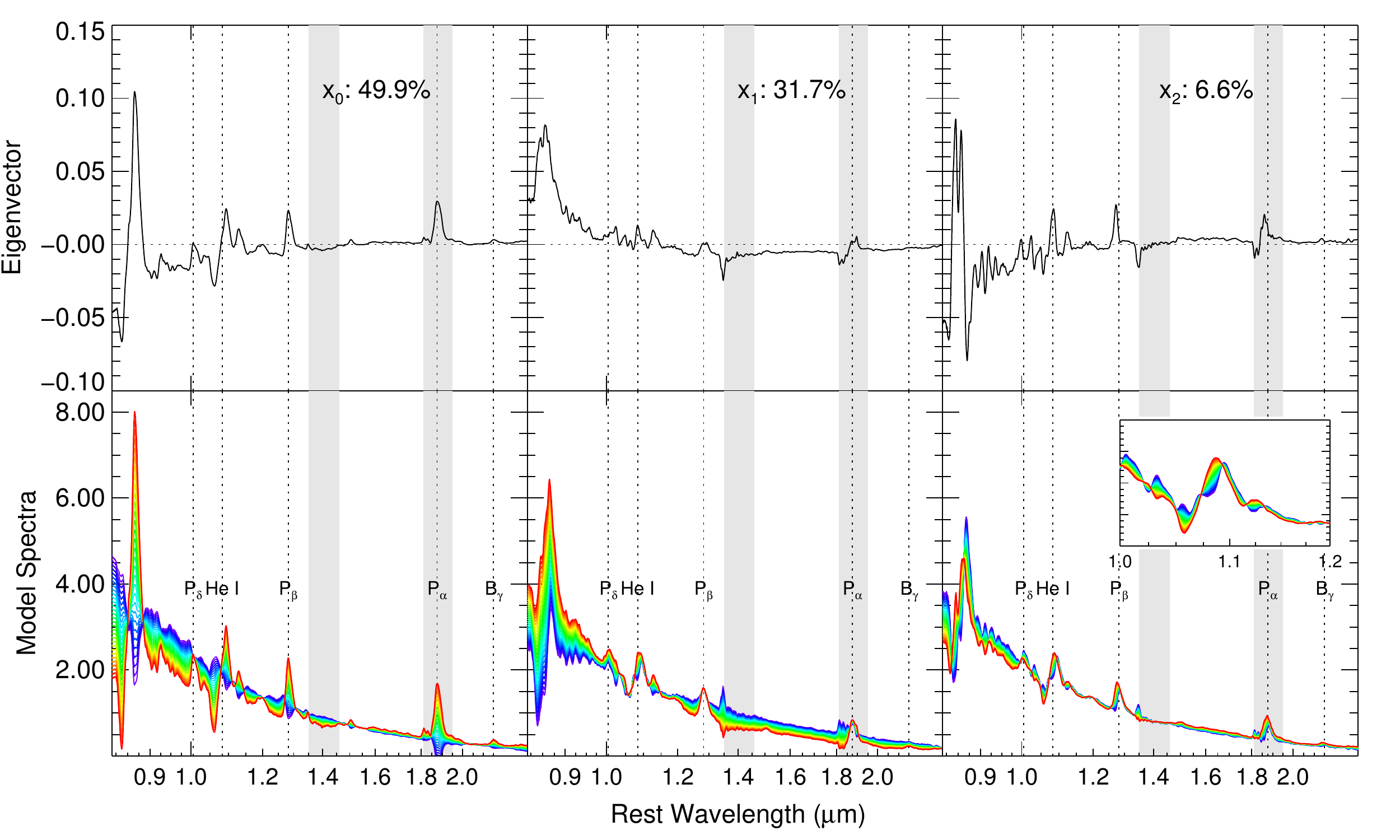}
    \caption{Principal component model of the SNe~II spectrum. The first three components are shown with the effects of $3\sigma$ variation. The eigenvalue for each projected component is noted on top as a percentage of the total variance. Rest wavelengths of the strongest features are plotted as vertical dashed lines: $P_{\alpha}$, $P_{\beta}$, He\,{\footnotesize I}, and $P_{\delta}$. The zeroth component represents the color and strength of features making up the spectrum over time. The first component shows calcium changes that vary with SN color. The second component picks out the different spectral classes, \emph{weak} and \emph{strong}. The bottom left panel also shows the zoomed in region from $1.0-1.2\,\mu$m where the spectral dichotomy is seen. The colors show deviations from the mean spectrum of $\pm3\sigma$.}
    \label{fig:pca02}
\end{figure*}

We require that each input spectrum for PCA have a S/N of at least 50 in the telluric regions, $1.35-1.45\,\mu$m and $1.80-2.00\,\mu$m, as PCA is susceptible to noise spikes. The zeroth principal component makes up $\sim$50$\%$ of the total variance of the spectrum, as shown in Figure \ref{fig:pca02}. This component describes the slope, line strength, and velocity over time of the SNe~II, suggesting these spectral properties are all correlated. There is a strong correlation between the projections of observed spectra onto the zeroth principle component and phase, as expected, since the line strengths of most features increase with time. The first component picks out $\sim31\%$ of the variance and shows further color and calcium changes when the region around He\,{\footnotesize I} $\lambda1.083\,\mu$m is uniform. The second component makes up $\sim6\%$ of the total variance and is of particular interest as it picks out the difference in this region between \emph{weak} and \emph{strong} SNe~II. The dichotomy of the region around He\,{\footnotesize I} $\lambda1.083\,\mu$m motivated splitting the sample into two templates and further supports the two distinct groups outlined in Section \ref{dichotomy}. 

\begin{figure*}
    \centering
    \includegraphics[width=1.5\columnwidth]{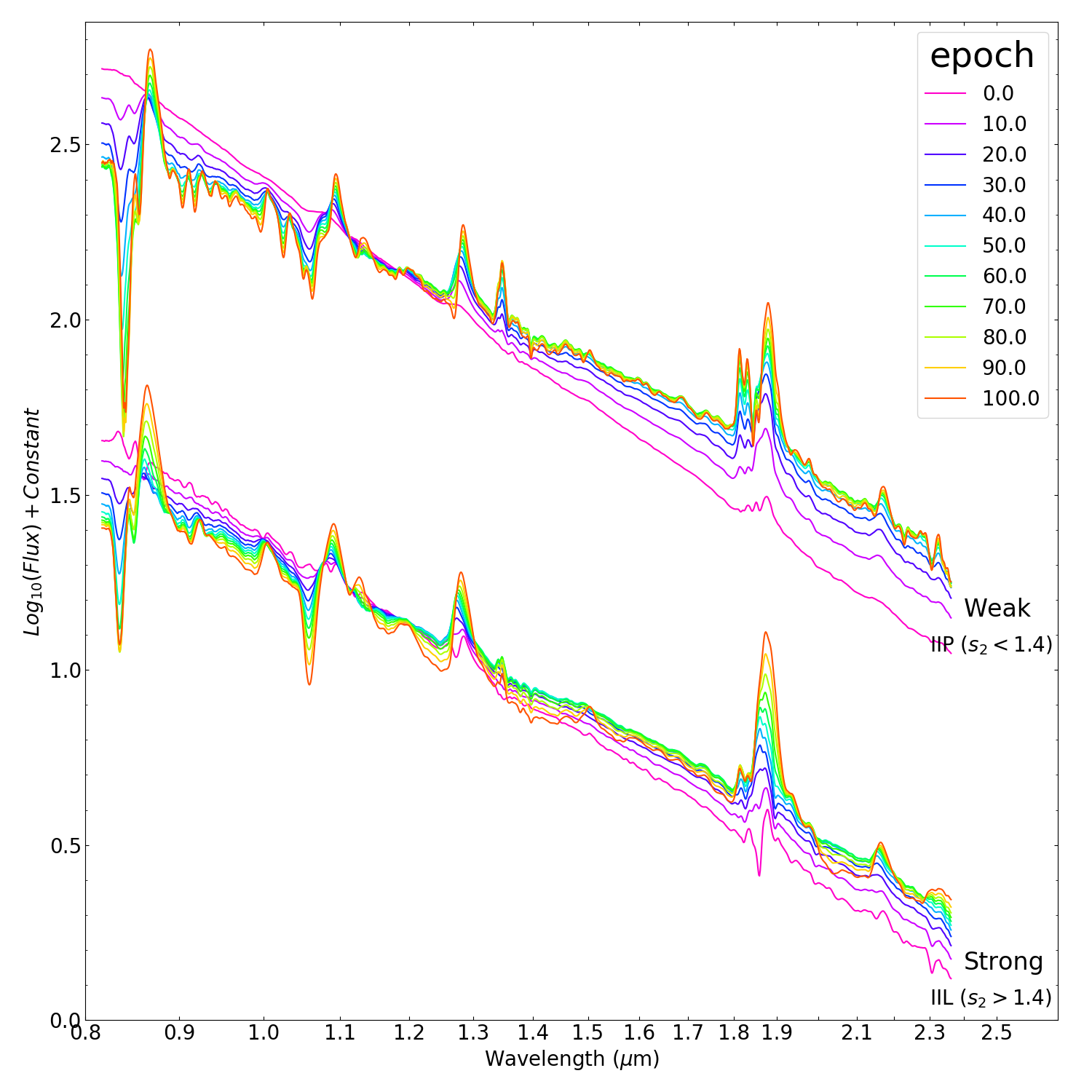}
    \caption{Time evolution of the two spectral templates; \emph{weak} and \emph{strong} SNe~II corresponding to slow and fast declining SNe~II (IIP and IIL, respectively). Different epochs are shown in color to illustrate the evolution over time of the models.}
    \label{fig:modeltime}
\end{figure*}

Cosmological studies using both Type II and Type Ia SNe are pushing into the NIR to minimize the effect of dust extinction \citep[e.g.][]{2019MNRAS.483.5459R}, and representative spectral templates of all types of SNe are crucial for future SN cosmological experiments. The two spectral templates are split by classification of \emph{weak} and \emph{strong} He\,{\footnotesize I} $\lambda1.083\,\mu$m absorption and are created by polynomial fitting to the projections over time. The resulting \emph{strong} and \emph{weak} templates are plotted, with different colors representing differents epochs, in Figure \ref{fig:modeltime}. We conclude that slow declining SNe~II (conventionally IIP) are best represented by the \emph{weak} spectral template while the \emph{strong} template better fits fast declining SNe~II (conventionally IIL). Our template spectra can be found on the Web.\footnote{https://csp.obs.carnegiescience.edu/data}

\section{Discussion} \label{discussion}

Due to the homogeneity of the hydrogen rich envelope, the difference in spectral features between fast and slow declining SNe~II most likely comes from explosion energy and temperature changes \citep{1988PASAu...7..434H,1990A&A...234..343T,1995A&A...297..802D,2000AJ....119.2968G}. The light curves are powered by the energy stored by the hydrodynamical shell during the early phase, the heating by radioactive decay of $^{56}Co \rightarrow ^{56}Ni \rightarrow ^{56}Fe$, and the recombination energy. During the plateau phase, the opacity drops by several orders of magnitude from the outer recombined layers above the photosphere to the inner ionized layers. The release of stored energy and the recombination of hydrogen powers the plateau phase \citep{2003MNRAS.345..111C}. The luminosity drops when the recombination front reaches the inner edge of the hydrogen-rich layers. The total energy released and the duration of the plateau decreases with the total hydrogen mass and, eventually, marks the transition from slow declining SNe~II with long plateaus to fast declining SN~II; however, explosion energy likely plays a significant role \citep{1993ApJ...414..712P}. With increasing hydrogen-mass, the C/O-rich layers are exposed later. This may explain why in SNe~II \emph{weak} we see later emission from the first overtone of the CO band \citep{2000AJ....119.2968G,2001RMxAC..10..195S,2017hsn..book.2125M,2018MNRAS.481..806B}, pointing to a mass sequence.

Progenitors of slow declining SNe~II have been interpreted to have lower zero-age main sequence mass than those of fast declining SNe~II, as they have retained much of their hydrogen envelope \citep{2000NewAR..44..297H}. Alternatively, \citet{2017ApJ...838...28M,2018ApJ...858...15M} suggest that SNe~IIL may be the result of a RSG surrounded by dense circumstellar material. If SNe~IIL have lost some of their hydrogen envelope, the process in which they lose this mass is uncertain. We cannot make a distinction in this work between SNe which have lost some of their hydrogen envelope by wind or a common envelope phase. 

The longer lifetime of presumably lower mass progenitors of the slower declining SNe~II leading up to the explosion allows more s-process elements, such as Sr, to be produced, providing a possible explanation for the Sr II lines observed mostly in the SNe~II \emph{weak} class \citep{2000ApJS..129..625L}. However, we see no correlation with spectral type and progenitor mass, determined using pre-explosion images, for those SNe within our sample and the sample of \cite{2015PASA...32...16S}, possibly due to the large uncertainty in mass.

The mixing of $^{56}Ni$ to high velocities could explain the pEW difference between SNe II \emph{weak} and \emph{strong}. The presence of He\,{\footnotesize I} requires high energy non-thermal photons close to the He layer \citet{1988ApJ...335L..53G}. This can only be achieved by gamma rays produced from the radioactive decay of $^{56}Ni$.

Multiple scenarios exist to explain the HV features in the spectral \emph{weak} class. The most likely explanation is from thermal excitation produced by a reverse shock as a result of the interaction between the SN ejecta and wind \citep{2007ApJ...662.1136C}, in which H and He are excited by X-rays, with the HV H possibly mixed into $H_{\alpha}$ and HV He present in the NIR. Alternatively, \cite{2008MNRAS.383...57D} suggest a radiation transport effect in layers with certain density slopes where hot, high-density inner layers and low-density outer layers with long recombination times are separated by a region of low excitation common to all SNe~II. However, this theory does not explain why these \emph{weak} features are not seen in fast declining SNe~II. Moreover, the models of \cite{2008MNRAS.383...57D} predict He\,{\footnotesize I} showing up first and vanishing after $\sim 20$ days, in contradiction to the observations. Thus we believe the HV features seen are powered by a shock.

\section{Conclusions} \label{conclusions}

We have presented 81 NIR spectra of 30 SNe~II observed between 2011 and 2015 as part of the CSP-II. The spectra range between 1 and 150 days post explosion. Using $V$-band light curves, photometric properties were measured for 14 SNe within the sample. The evolution of NIR spectral features was outlined and for the most dominant features, pEW and velocities were systematically measured. 

There is a strong dichotomy in the NIR spectral features of SNe~II, and it is the most prominent in the features around He\,{\footnotesize I} $\lambda1.083\,\mu$m. Thus, we presented spectral classifications based on the strength of this feature: \emph{strong} and \emph{weak} SNe~II. Characteristics of the two classes are outlined as follows:
\begin{itemize}
    \item \emph{Weak} SNe show an accompanying absorption feature on the blue side of the He\,{\footnotesize I} $\lambda1.083\,\mu$m, feature A, which we interpret as a HV component of the same He\,{\footnotesize I} line, and \emph{strong} SNe do not. The HV He\,{\footnotesize I} line always has an earlier onset then the photospheric He\,{\footnotesize I} $\lambda1.083\,\mu$m absorption feature.
    \item \emph{Weak} SNe show earlier absorption, $\sim20$ days past explosion, from $P_{\gamma}$/Sr\,{\footnotesize II} than \emph{strong} SNe. \emph{Strong} SNe show this feature at $\sim40$ days.
    \item \emph{Weak} SNe more often exhibit Sr\,{\footnotesize II} features in the $1.0-1.1\,\mu$m region.
    \item \emph{Strong} SNe can form the first overtone of CO at earlier times, less than $100$ days. \emph{Weak} SNe have later formation of CO, past 120 days.
\end{itemize}

We found that these spectral classifications of \emph{strong} and \emph{weak} SNe correspond to the plateau decline rate: slow declining SNe~II (IIP) are \emph{weak} and fast declining SNe~II (IIL) are \emph{strong}. This a somewhat surprising result given the previous research showing a continuum in optical light curves and spectroscopic features.
The HV He\,{\footnotesize I} feature seen is most likely due to a reverse shock as explained in \cite{2007ApJ...662.1136C}. However, this does not explain the dichotomy. PCA was performed on the spectra which further confirmed the observed dichotomy that represented $\sim6$\% of the spectroscopic variance.  Finally, using PCA, two spectral templates for SNe~II were created, SNe \emph{weak} and \emph{strong}. These templates are crucial for cosmological studies using SNe~II.

\acknowledgments
The authors would like to thank the anonymous referee for their comments. The work of the CSP-II has been generously supported by the National Science Foundation (NSF) under grants AST-1008343, AST-1613426, AST-1613455, and AST-1613472. The CSP-II was also supported in part by the Danish Agency for Science and Technology and Innovation through a Sapere Aude Level 2 grant. 

We would like to thank Michael Cushing for his work on Spextool in processing low-resolution data. PH acknowledges support by the NSF grant 1715133. Research by DJS is supported by NSF grants AST-1821987, AST-1821967, AST-1813708, and AST-1813466. CPG acknowledges support from EU/FP7-ERC grant no. [615929]. Support for JLP is provided in part by FONDECYT through the grant 1191038 and by the Ministry of Economy, Development, and Tourism’s Millennium Science Initiative through grant IC120009, awarded to The Millennium Institute of Astrophysics, MAS.

\software{firehose \citep{2013PASP..125..270S}, xtellcor \citep{2003PASP..115..389V}, Spextool \citep{2004PASP..116..362C}, SNID \citep{2007ApJ...666.1024B}}


\begin{deluxetable*}{ccccccccccc}
\addtolength{\tabcolsep}{-1.5pt}
\rotate
\tablecaption{Supernovae in the Sample}
\tablenum{1}
\label{tab:sample}
\tablecolumns{10}
\tablehead{
\colhead{Name} &
\colhead{Type} &
\colhead{$s_2$} &
\colhead{NIR} &
\colhead{$V$-band} &
\colhead{Discovery} &
\colhead{Discovery} &
\colhead{Last Non-Detection} &
\colhead{Discovery} &
\colhead{Last Non-Detection} \\
\colhead{} &
\colhead{} &
\colhead{(mag 100 day\textsuperscript{-1})} &
\colhead{Spectra} &
\colhead{Photometry} &
\colhead{(YYYY-MM-DD)} &
\colhead{(MJD)} &
\colhead{(MJD)} &
\colhead{Reference} &
\colhead{Reference}
}
\startdata
ASASSN-13dn & II & - & 1 & No & 2013-12-11.6 & 56637.6 & 56497.3 & \cite{2013ATel.5665....1S} & \cite{2013ATel.5665....1S} \\
ASASSN-14gm & IIP & 0.43 (0.36) & 4 & Yes & 2014-09-02.5 & 56902.5 & 56899.5 & \cite{14gmDisc} & \cite{2016MNRAS.459.3939V} \\
ASASSN-14jb & IIP & -0.03 (0.14) & 1 & Yes & 2014-10-19.1 & 56949.1 & 56943.0 & \cite{2014ATel.6592....1B} & \cite{2014ATel.6592....1B} \\
ASASSN-15bb & IIP & 0.47 (0.06) & 5 & Yes & 2015-01-16.3 & 57038.3 & 57036.3 & \cite{2015ATel.6936....1K} & \cite{2015ATel.6936....1K} \\
ASASSN-15fz & IIP & 0.94 (0.19) & 2 & Yes & 2015-03-30.6 & 57111.6 & 57110.3 & \cite{2015ATel.7320....1B} & \cite{2015ATel.7320....1B} \\
ASASSN-15jp & II & - & 1 & Yes & 2015-05-21.1 & 57163.1 & 57161.0 & \cite{2015ATel.7546....1H} & \cite{2015ATel.7546....1H} \\
CATA13A & II & - & 1 & No & 2013-12-07.2 & 56633.2 & 56415.9 & \cite{13ADisc} & \cite{2013CBET.3752....1M} \\
KISS14J & II & - & 1 & No & 2014-02-23.5 & 56711.5 & - & \cite{14jDisc} & - \\
LSQ12bri & II & - & 1 & No & 2012-04-06 & 56023 & - & \cite{12briDisc} & - \\
LSQ12dcl & II & - & 1 & No & 2012-06-24.4 & 56102.4 & - & \cite{12dclDisc} & - \\
LSQ13dpa & IIP & 0.28 (2.01) & 1 & Yes & 2013-12-18.3 & 56644.3 & - & \cite{13dpaDisc} & - \\
LSQ15ok & IIP & 1.03 (0.11) & 1 & Yes & 2015-02-02 & 57069.6 & 57055 & \cite{15okDisc} & \cite{2015ATel.7143....1Z}\\
iPTF13dqy & II & - & 1 & No & 2013-10-07.5 & 56572.5 & 56570.4 & \cite{13dqyDisc} & \cite{2017NatPh..13..510Y} \\
iPFT13dzb & II & - & 1 & No & 2013-11-08 & 56604 & - & \cite{13dzbDisc} & - \\
SN~2012A & IIP & 0.94 (0.09) & 1 & Yes & 2012-01-07.4 & 55933.4 & 55924 & \cite{12ADisc} & \cite{2012CBET.2974....2L}\\
SN~2012aw & IIP & 1.27 (0.07) & 5 & Yes & 2012-03-16.9 & 56002.9 & 56001.9 & \cite{12awDisc} & \cite{2014ApJ...787..139D} \\
SN~2012hs & II & - & 1 & No & 2012-12-15.2 & 56276.2 & 56272.3 & \cite{2012CBET.3347....1C} & \cite{2012CBET.3347....1C} \\
SN~2013ab & IIP & 1.33 (0.01) & 4 & Yes & 2013-02-17.5 & 56340.5 & 56338 & \cite{2013ATel.4823....1Z} & \cite{2013ATel.4823....1Z}\\
SN~2013ai & IIL & 1.61 (0.01) & 1 & Yes & 2013-03-01.7 & 56352.7 & 56329.7 & \cite{2013ATel.4849....1C} & \cite{2013ATel.4849....1C} \\
SN~2013by & IIL & 2.8 (0.01) & 3 & Yes & 2013-04-24.3 & 56406.3 & - & \cite{13byDisc} & - \\
SN~2013ej & IIL & 2.34 (0.79) & 5 & Yes & 2013-07-25.5 & 56498.5 & - & \cite{13ejDisc} & - \\
SN~2013gd & IIP & 0.64 (0.02) & 12 & Yes & 2013-11-09.4 & 56605.4 & 56601 & \cite{2013ATel.5563....1Z} & \cite{2013ATel.5563....1Z} \\
SN~2013gu & IIL & 1.91 (0.04) & 6 & Yes & 2013-12-05.8 & 56631.8 & 56212.9 & \cite{2013ATel.5630....1S} & \cite{2013ATel.5630....1S} \\
SN~2013hj & IIL & 1.59 (0.66) & 14 & Yes & 2013-12-12.3 & 56638.3 & - & \cite{13hjDisc} & - \\
SN~2013ht & II & - & 1 & No & 2013-12-31.4 & 56657.4 & - & \cite{13htDisc} & - \\
SN~2014A & II & - & 1 & No & 2014-01-01.6 & 56658.6 & - & \cite{14ADisc} & - \\
SN~2014bt & II & - & 2 & No & 2014-05-31.6 & 56809.6 & - & \cite{14btDisc} & - \\
SN~2014cw & IIP & 0.54 (0.09) & 1 & Yes & 2014-08-29.5 & 56898.5 & 56842.9 & \cite{2014ATel.6435....1S} & \cite{2014ATel.6435....1S} \\ 
SN~2014cy & II & - & 1 & No & 2014-08-31.7 & 56900.7 & - & \cite{14cyDisc} & - \\
SN~2014dw & IIL & 2.84 (0.07) & 1 & Yes & 2014-11-6.6 & 56967.6 & - & \cite{14dwDisc} & - \\
\enddata
\tablecomments{Some discovery and last non-detection observations were reported without time of day and thus are quoted here as such.}
\end{deluxetable*}

\clearpage
\startlongtable
\begin{deluxetable*}{cccccccc}
\tablecaption{Log of NIR observations}
\tablenum{2}
\tablecolumns{8}
\tablewidth{0pt}
\tablehead{
\colhead{SN} &
\colhead{UT Date} &
\colhead{JD} &
\colhead{Phase} &
\colhead{Instrument} &
\colhead{Telescope} &
\colhead{S/N}
}
\startdata
ASASSN-13dn & 2014-02-22 & 2456711 & 143.5 & FIRE & Baade & 113\\
ASASSN-14gm & 2014-09-03 & 2456904 & 0.0 & FIRE & Baade & 129\\
 & 2014-09-23 & 2456924 & 21.0 & SpeX & IRTF & 99\\
 & 2014-12-14 & 2457006 & 102.0 & FIRE & Baade & 101\\
 & 2015-01-05 & 2457028 & 124.0 & FIRE & Baade & 35\\
ASASSN-14jb & 2014-11-05 & 2456967 & 20.0 & FIRE & Baade & 111\\
ASASSN-15bb & 2015-01-28 & 2457051 & 13.8 & FIRE & Baade & 375\\
 & 2015-03-07 & 2457089 & 51.8 & FIRE & Baade & 109\\
 & 2015-04-07 & 2457120 & 82.8 & FIRE & Baade & 109\\
 & 2015-04-12 & 2457125 & 87.8 & FIRE & Baade & 53\\
 & 2015-06-02 & 2457176 & 138.8 & FIRE & Baade & 43\\
ASASSN-15fz & 2015-04-02 & 2457115 & 4.0 & FIRE & Baade & 142\\
 & 2015-04-07 & 2457120 & 9.0 & FIRE & Baade & 155\\
ASASSN-15jp & 2015-06-02 & 2457176 & 14.0 & FIRE & Baade & 310\\
CATA13A & 2013-12-09 & 2456636 & 111.5 & FIRE & Baade & 71\\
KISS14J & 2014-02-27 & 2456716 & 15.5 & FIRE & Baade & 44\\
LSQ12bri & 2012-04-08 & 2456026 & 3.0 & FIRE & Baade & 107\\
LSQ12dcl & 2012-06-26 & 2456105 & 2.5 & FIRE & Baade & 69\\
LSQ13dpa & 2013-12-20 & 2456647 & 4.8 & FIRE & Baade & 60\\
LSQ15ok & 2015-03-07 & 2457089 & 17.0 & FIRE & Baade & 102\\
iPTF13dqy & 2013-11-14 & 2456611 & 39.5 & FIRE & Baade & 69\\
iPTF13dzb & 2013-11-20 & 2456617 & 13.0 & FIRE & Baade & 54\\
SN~2012A & 2012-01-15 & 2455942 & 13.2 & FIRE & Baade & 589\\
SN~2012aw & 2012-04-08 & 2456026 & 23.5 & FIRE & Baade & 155\\
 & 2012-04-11 & 2456029 & 26.5 & FIRE & Baade & 259\\
 & 2012-04-19 & 2456037 & 34.5 & FIRE & Baade & 173\\
 & 2012-04-30 & 2456048 & 45.5 & FIRE & Baade & 135\\
 & 2012-05-07 & 2456055 & 52.5 & FIRE & Baade & 124\\
SN~2012hs & 2013-01-06 & 2456299 & 24.8 & FIRE & Baade & 56\\
SN~2013ab & 2013-02-28 & 2456352 & 12.8 & FIRE & Baade & 172\\
 & 2013-03-20 & 2456372 & 32.8 & FIRE & Baade & 186\\
 & 2013-05-19 & 2456432 & 92.8 & FIRE & Baade & 92\\
SN~2013ai & 2013-03-20 & 2456372 & 24.0 & FIRE & Baade & 150\\
SN~2013by & 2013-04-24 & 2456407 & 3.0 & FIRE & Baade & 318\\
 & 2013-05-22 & 2456435 & 31.0 & FIRE & Baade & 121\\
 & 2013-07-29 & 2456503 & 99.0 & FIRE & Baade & 54\\
SN~2013ej & 2013-07-29 & 2456503 & 5.8 & FIRE & Baade & 573\\
 & 2013-09-11 & 2456547 & 49.8 & SpeX & IRTF & 231 \\
 & 2013-10-25 & 2456591 & 93.8 & FIRE & Baade & 50\\
 & 2013-11-14 & 2456611 & 113.8 & FIRE & Baade & 29\\
 & 2013-12-14 & 2456641 & 143.8 & FIRE & Baade & 96\\
SN~2013gd & 2013-11-14 & 2456611 & 7.8 & FIRE & Baade & 105\\
 & 2013-11-30 & 2456627 & 23.8 & FIRE & Baade & 283\\
 & 2013-12-09 & 2456636 & 32.8 & FIRE & Baade & 102\\
 & 2013-12-14 & 2456641 & 37.8 & FIRE & Baade & 110\\
 & 2013-12-20 & 2456647 & 43.8 & FIRE & Baade & 139\\
 & 2013-12-27 & 2456654 & 50.8 & FIRE & Baade & 57\\
 & 2014-01-01 & 2456659 & 55.8 & FIRE & Baade & 182\\
 & 2014-01-09 & 2456667 & 63.8 & FIRE & Baade & 76\\
 & 2014-01-14 & 2456672 & 68.8 & FIRE & Baade & 41\\
 & 2014-02-08 & 2456697 & 93.8 & FIRE & Baade & 45\\
 & 2014-02-27 & 2456716 & 112.8 & FIRE & Baade & 68\\
 & 2014-03-18 & 2456735 & 131.8 & FIRE & Baade & 27\\
SN~2013gu & 2013-12-09 & 2456636 & 3.5 & FIRE & Baade & 234\\
 & 2013-12-14 & 2456641 & 8.5 & FIRE & Baade & 90\\
 & 2013-12-20 & 2456647 & 14.5 & FIRE & Baade & 96\\
 & 2013-12-27 & 2456654 & 21.5 & FIRE & Baade & 84\\
 & 2014-01-09 & 2456667 & 34.5 & FIRE & Baade & 68\\
 & 2014-01-14 & 2456672 & 39.5 & FIRE & Baade & 42\\
SN~2013hj & 2013-12-14 & 2456641 & 4.0 & FIRE & Baade & 305\\
 & 2013-12-20 & 2456647 & 10.0 & FIRE & Baade & 366\\
 & 2013-12-27 & 2456654 & 17.0 & FIRE & Baade & 219\\
 & 2014-01-01 & 2456659 & 22.0 & FIRE & Baade & 205\\
 & 2014-01-09 & 2456667 & 30.0 & FIRE & Baade & 93\\
 & 2014-02-08 & 2456697 & 60.0 & FIRE & Baade & 267\\
 & 2014-02-15 & 2456704 & 67.0 & FIRE & Baade & 107\\
 & 2014-02-22 & 2456711 & 74.0 & FIRE & Baade & 161\\
 & 2014-02-27 & 2456716 & 79.0 & FIRE & Baade & 95\\
 & 2014-03-10 & 2456727 & 90.0 & FIRE & Baade & 154\\
 & 2014-03-25 & 2456742 & 105.0 & FIRE & Baade & 64\\
 & 2014-04-23 & 2456771 & 134.0 & FIRE & Baade & 39\\
SN~2013ht & 2014-01-01 & 2456659 & 1.5 & FIRE & Baade & 62\\
SN~2014A & 2014-01-02 & 2456660 & 45.5 & FIRE & Baade & 77\\
SN~2014bt & 2014-06-06 & 2456815 & 5.5 & FIRE & Baade & 226\\
 & 2014-07-10 & 2456849 & 39.5 & FIRE & Baade & 334\\
SN~2014cw & 2014-09-03 & 2456904 & 33.2 & FIRE & Baade & 130\\
SN~2014cy & 2014-09-03 & 2456904 & 12.0 & FIRE & Baade & 119\\
SN~2014dw & 2014-12-14 & 2457006 & 38.5 & FIRE & Baade & 65\\
\enddata
\label{tab:NIR}
\end{deluxetable*}

\begin{deluxetable*}{cccccc}
\tablecaption{Supernova Host Galaxies}
\tablenum{3}
\label{tab:hosts}
\tablecolumns{6}
\tablewidth{0pt}
\tablehead{
\colhead{Name} &
\colhead{Redshift} &
\colhead{RA} &
\colhead{Dec} &
\colhead{Host Galaxy}
}
\startdata
ASASSN-13dn & 0.023 & 12:52:58.39 & $+$32:25:05.60 & SDSS J125258.03+322444.3\\
ASASSN-14gm & 0.006 & 00:59:47.83 & $-$07:34:19.30 & NGC 337\\
ASASSN-14jb & 0.006 & 22:23:16.12 & $-$28:58:30.78 & ESO 467-G051\\
ASASSN-15bb & 0.016 & 13:01:06.38 & $-$36:36:00.17 & ESO 381-IG048\\
ASASSN-15fz & 0.017 & 13:35:25.14 & $+$01:24:33.00 & NGC 5227\\
ASASSN-15jp & 0.010 & 10:11:38.99 & $-$31:39:04.04 & NGC 3157\\
CATA13A & 0.035 & 06:25:10.07 & $-$37:20:41.30 & ESO 365-G16\\
KISS14J & 0.018 & 11:14:52.16 & $+$19:27:17.80 & NGC 3859\\
LSQ12bri & 0.030 & 13:35:48.35 & $-$21:23:53.47 & Unknown \\
LSQ12dcl & 0.031 & 00:13:43.35 & $-$00:27:58.38 & SDSS J001343.81-002735.7\\
LSQ13dpa & 0.024 & 11:01:12.91 & $-$5:50:52.57 & LCSB S1492O\\
LSQ15ok & 0.014 & 10:49:16.67 & $-$19:38:26.01 & ESO 569-G12\\
iPTF13dqy & 0.012 & 23:19:44.70 & $+$10:11:04.40 & NGC 7610\\
iPFT13dzb & 0.037 & 03:10:50.20 & $-$00:21:40.30 & 2MASX J03104933-0021256\\
SN~2012A & 0.003 & 10:25:07.38 & $+$17:09:14.60 & NGC 3239\\
SN~2012aw & 0.003 & 10:43:53.72 & $+$11:40:17.70 & M95\\
SN~2012hs & 0.006 & 09:49:14.71 & $-$47:54:45.60 & ESO 213-2\\
SN~2013ab & 0.005 & 14:32:44.49 & $+$09:53:12.30 & NGC 5669\\
SN~2013ai & 0.009 & 06:16:18.35 & $-$21:22:32.90 & NGC 2207\\
SN~2013by & 0.004 & 16:59:02.43 & $-$60:11:41.80 & ESO 138-G10\\
SN~2013ej & 0.002 & 01:36:48.16 & $+$15:45:31.00 & M74\\
SN~2013gd & 0.013 & 03:49:05.64 & $-$03:03:28.30 & MCG-01-10-039\\
SN~2013gu & 0.018 & 01:46:38.27 & $+$04:13:24.40 & SDSS J014638.24+041333.3\\
SN~2013hj & 0.007 & 09:12:06.29 & $-$15:25:46.00 & MCG -02-24-003\\
SN~2013ht & 0.028 & 10:55:50.95 & $-$09:51:42.40 & MCG -02-28-21\\
SN~2014A & 0.006 & 13:16:59.36 & $-$16:37:57.00 & NGC 5054\\
SN~2014bt & 0.016 & 21:43:11.13 & $-$38:58:05.80 & IC 5128\\
SN~2014cw & 0.009 & 22:15:26.55 & $-$10:28:34.60 & PGC 68414\\ 
SN~2014cy & 0.006 & 23:44:16.03 & $+$10:46:12.50 & NGC 7742\\
SN~2014dw & 0.008 & 11:10:48.41 & $-$37:27:02.20 & NGC 3568\\
\enddata
\end{deluxetable*}

\begin{deluxetable*}{ccccc}
\tablecaption{Supernova Distances}
\tablenum{4}
\label{tab:distances}
\tablecolumns{5}
\tablewidth{0pt}
\tablehead{
\colhead{SN} &
\colhead{Redshift} &
\colhead{Redshift Independent} &
\colhead{Method} & 
\colhead{Reference}
\\
\colhead{} &
\colhead{Distance Modulus} &
\colhead{Distance Modulus} &
\colhead{} & 
\colhead{}
}
\startdata
ASASSN-13dn & 34.98 (0.09) & - & - & - \\
ASASSN-14gm & 31.63 (0.43) & 31.38 (0.20) & Tully-Fisher & \cite{2013AJ....146...86T} \\
ASASSN-14jb & 32.03 (0.36) & 30.89 (0.32) & Tully-Fisher & \cite{1992ApJS...81..413M} \\
ASASSN-15bb & 34.18 (0.14) & - & - & -\\
ASASSN-15fz & 34.31 (0.13) & - & - & -\\
ASASSN-15jp & 32.91 (0.24) & 33.26 (0.13) & Tully-Fisher & \cite{2014MNRAS.445.2677S} \\
CATA13A & 35.91 (0.06) & - & - & -\\
KISS14J & 34.43 (0.12) & 35.03 (0.08) & Tully-Fisher & \cite{2014MNRAS.445.2677S} \\
LSQ12bri & 35.56 (0.07) & - & - & -\\
LSQ12dcl & 35.64 (0.07) & - & - & -\\
LSQ13dpa & 34.98 (0.09) & - & - & -\\
LSQ15ok & 33.88 (0.16) & - & - & -\\
iPTF13dqy & 33.54 (0.18) & 33.51 (0.10) & Tully-Fisher & \cite{2014MNRAS.445.2677S} \\
iPFT13dzb & 36.03 (0.06) & 37.35 (0.94) & FP & \cite{2014MNRAS.445.2677S} \\
SN~2012A & 30.52 (0.72) & - & - & - \\
SN~2012aw & 30.00 (0.09) & 30.10 (0.06) & Cepheids & \cite{2001ApJ...553...47F} \\
SN~2012hs & 32.03 (0.36) & - & - & -\\
SN~2013ab & 31.63 (0.43) & - & - & - \\
SN~2013ai & 32.91 (0.24) & 32.99 (0.30) & SNIa & \cite{1982ASIC...90..221A} \\
SN~2013by & 31.15 (0.54) & 30.84 (0.80) & Tully-Fisher & \cite{1988cng..book.....T} \\
SN~2013ej & 29.64 (1.09) & 30.04 (0.03) & TRGB & \cite{2014ApJ...792...52J} \\
SN~2013gd & 33.72 (0.17) & - & - & -\\
SN~2013gu & 34.43 (0.12) & - & - & -\\
SN~2013hj & 32.37 (0.31) & 31.94 (0.80) & Tully-Fisher & \citet{Makarov} \\
SN~2013ht & 35.41 (0.07) & - & - & -\\
SN~2014A & 32.03 (0.36) & 31.30 (0.20) & Tully-Fisher & \cite{2013AJ....146...86T} \\
SN~2014bt & 34.15 (0.14) & - & - & -\\
SN~2014cw & 32.91 (0.24) & - & - & -\\
SN~2014cy & 32.03 (0.36) & 31.73 (0.80) & Tully-Fisher & \cite{1988cng..book.....T} \\
SN~2014dw & 32.66 (0.27) & 32.63 (0.10) & Sosies & \citet{Terry}
\enddata
\tablecomments{Assuming $H_{0}=71,\Omega_{m}=0.27, \Omega_{}=0.73$, and that each has a peculiar velocity of $300$ km s\textsuperscript{-1}. The distance modulus error is given in parenthesis.}
\end{deluxetable*}

\begin{deluxetable*}{cccccc}
\tablecaption{Previously published supernova classifications.}
\tablenum{5}
\tablecolumns{5}
\label{tab:historical}
\tablewidth{0pt}
\tablehead{
\colhead{Name} &
\colhead{Photometric} &
\colhead{$s_2$} &
\colhead{NIR Spectroscopic} & 
\colhead{Publication}
\\
\colhead{} & 
\colhead{Classification} & 
\colhead{(mag 100 days\textsuperscript{-1})} &
\colhead{Classification} & 
\colhead{}
}
\startdata
SN~1997D & - & - & Weak & \cite{2001MNRAS.322..361B} \\
SN~1999em & IIP & 0.37 (0.02) & - & \cite{2001ApJ...558..615H}, \cite{2003MNRAS.338..939E} \\
SN~2002hh & IIP & 0.73 (0.18) & - & \cite{2006MNRAS.368.1169P} \\
SN~2004et & IIP & 0.93 (0.15) & - & \cite{2010MNRAS.404..981M} \\
SN~2005cs & IIP & 0.43 (0.12) & Weak & \cite{2009MNRAS.394.2266P} \\
SN~2008in & IIP & 0.89 (0.03) & Weak & \cite{2014MNRAS.438..368T} \\
SN~2009N & IIP & 0.90 (0.22) & Weak & \cite{2014MNRAS.438..368T} \\
SN~2009ib & IIP & 0.27 (0.07) & Weak & \cite{2015MNRAS.450.3137T} \\
SN~2009md & IIP & - & Weak & \cite{2011MNRAS.417.1417F} \\
SN~2012A & IIP & 0.94 (0.09) & Strong & \cite{2013MNRAS.434.1636T} \\
SN~2012aw & IIP & 1.27 (0.07) & Weak & \cite{2014ApJ...787..139D}, \cite{2014MNRAS.439.3694J} \\
SN~2013by & IIL & 2.80 (0.01) & Strong & \cite{2015MNRAS.448.2608V} \\
SN~2013ej & IIL & 2.34 (0.79) & Strong & \cite{2014MNRAS.438L.101V} \\
ASASSN-15oz & IIL & - & Strong & \cite{2019MNRAS.485.5120B} \\
SN~2016ija & IIL & 2.39 (0.38) & Strong & \cite{2018ApJ...853...62T} \\
SN~2017eaw & IIP & 0.74 (0.15) & Weak & \cite{2019AAS...23333502R} \\
SN~2017gmr & IIP & - & Weak & \cite{2019arXiv190701013A}
\enddata
\tablecomments{SN~2008in, SN~2009md, and SN~2012A were spectroscopically classified based on the features present in their published spectra, not a pEW measurement, as their data is not publicly accessible. A value of $s_2$ cannot be measured for SN~2009md and ASASSN-15oz as they have no publicly available photometry, thus their photometric classification is taken from their respective publications. SN~1997D has sparse photometric coverage and an uncertain explosion date so no value of $s_2$ can be measured.}
\end{deluxetable*}

\appendix

\section{Light Curves} \label{appendix}
\renewcommand\thefigure{\thesection.\arabic{figure}}
\renewcommand{\thetable}{\thesection.\arabic{table}}
\setcounter{figure}{0}
\setcounter{table}{0}

\begin{figure*}
    \includegraphics[width=\textwidth]{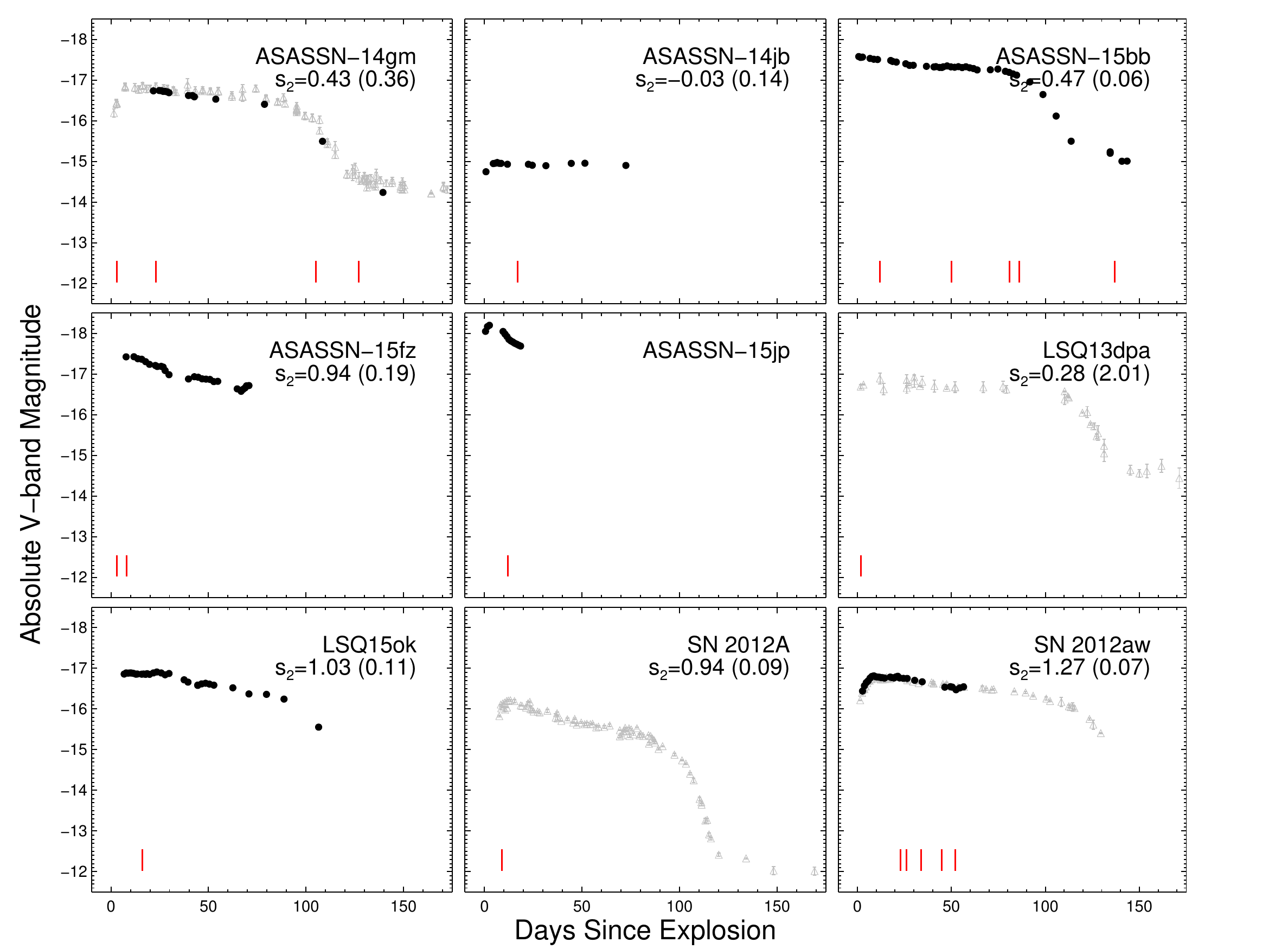}
    \caption{Milky Way extinction corrected $V$-band light curves that were measured. Black dots are data from CSP-II, and grey triangles represent previously published data. Red ticks at the bottom of each panel represent the epoch of NIR spectra taken. The optical data is taken from previously published data.}
    \label{fig:vSplit1}
\end{figure*}

\begin{figure*}
    \includegraphics[width=\textwidth]{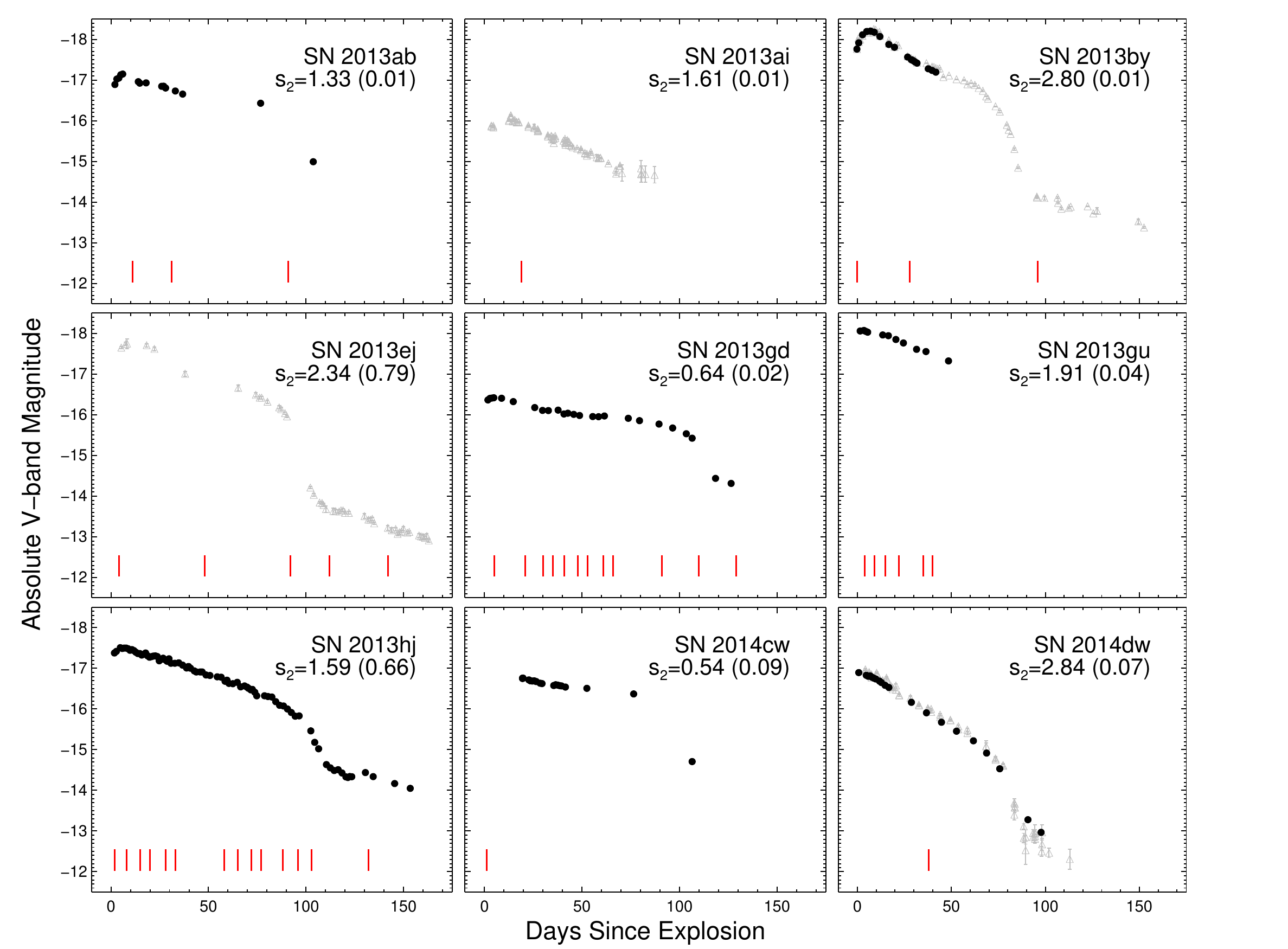}
    \caption{Figure \ref{fig:vSplit1} continued.}
    \label{fig:vSplit2}
\end{figure*}

For early epochs and plateau phase, a four-parameter fit is applied to determine the decline rates ($s_{1}$ and $s_{2}$), the transition epoch between the two decline rates ($t_{\rm trans}$), and the magnitude offset of the light curve. Fitting is also performed with multiple linear three parameter functions in order to determine if the light curve has two distinct early time decline phases characterized by multiple slopes. Bayesian Information Criterion \citep{1978AnSta...6..461S} was used to determine the goodness of fit between the three and four-parameter models. A linear fit is also applied to determine the radioactive decay tail slope, $s_{3}$, if sufficient photometric data past the plateau exists. The midpoint time of transition, $t_{\rm PT}$ between the plateau and linear decline phases is obtained through fitting the light curves with an eight-parameter fit function from \cite{2010ApJ...715..833O},
 \begin{equation}
     F(t) = \frac{-a_0}{1+e^{(t-t_{PT})/w_0}}+p_0(t-t_{PT})+m_0-Pe^{-(\frac{t-Q}{R})^2},
 \end{equation}
 where $t_{\rm PT}$ corresponds to the midpoint of transition between plateau and radioactive tail, and $m_0$ is the zero point magnitude at the transition time. The first term is a Fermi-Dirac function that describes the transition between plateau and linear decline phases. The second term and zero point magnitude, $m_0$, describe the radioactive tail of the light curve and offset the fit, respectively. The last term is a Gaussian that describes the shape of the light curve during the plateau. Figures \ref{fig:vSplit1} and \ref{fig:vSplit2} show the  CSP-II $V$-band light curves, as well as other light curves from the literature. Results of the fitting are listed in Table \ref{appendix:LCparams}.

\begin{figure*}
    \includegraphics[width=\textwidth]{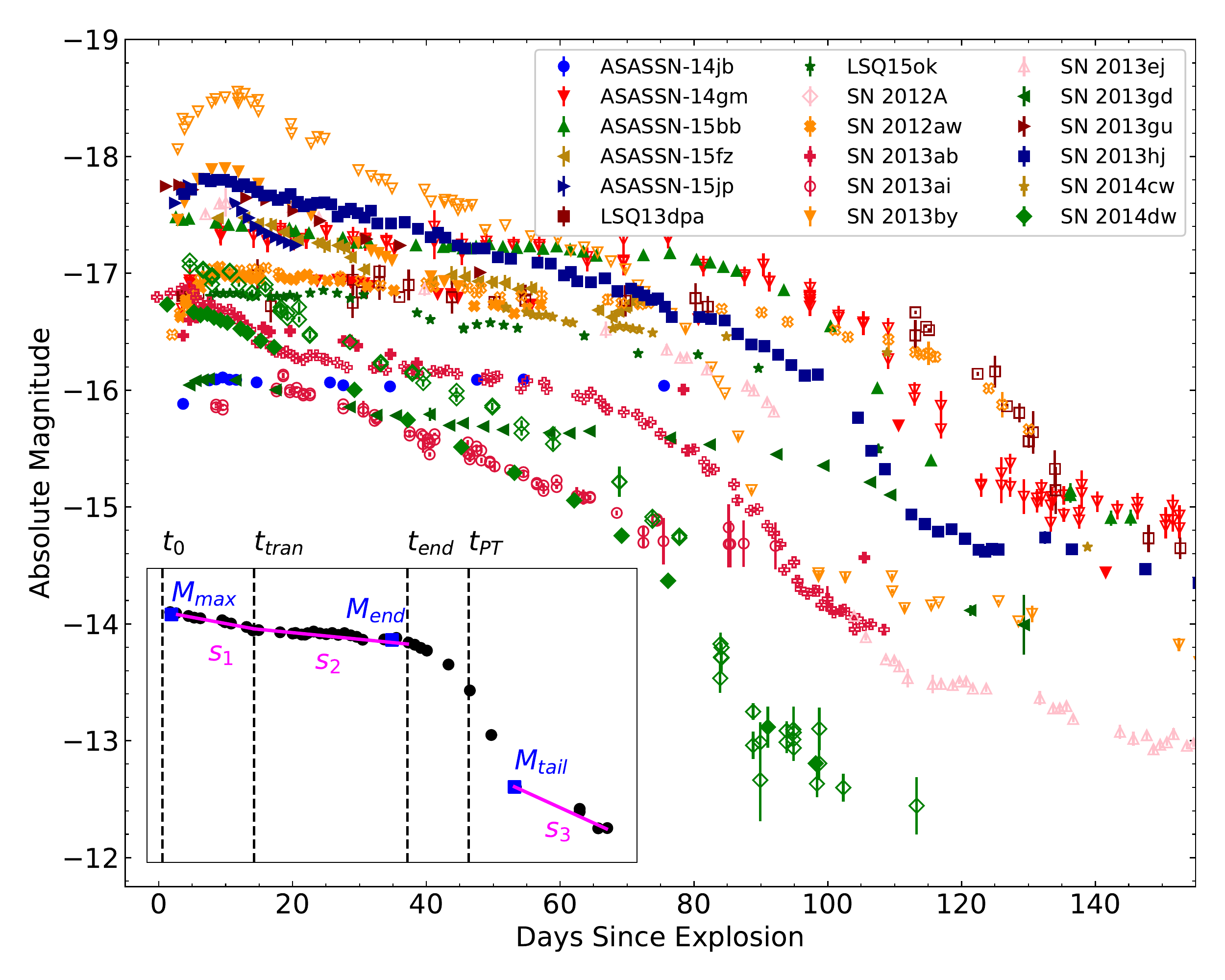}
    \caption{All Milky Way extinction corrected $V$-band light curve data plotted together in absolute magnitude. Errors are shown for each data point, unless smaller than the symbol. Open symbols represent data gathered from previously published work. The inset shows an example light fit with the parameters outlined in the text.}
    \label{fig:vPhot}
\end{figure*}


\setcounter{figure}{1}
\setcounter{table}{0}
\begin{deluxetable*}{cccccccccccc}
\addtolength{\tabcolsep}{-2.5pt}
\rotate
\tabletypesize{\scriptsize}
\tablecaption{$V$-band light curve measurements}
\label{appendix:LCparams}
\tablecolumns{10}
\tablewidth{0pt}
\tablehead{
\colhead{SN} & 
\colhead{$s_{\rm 1}$} & 
\colhead{$s_{\rm 2}$} & 
\colhead{$s_{\rm 3}$} & 
\colhead{$M_{\rm max}$} & 
\colhead{$M_{\rm end}$} &
\colhead{$M_{\rm tail}$} &
\colhead{$t_{\rm tran}$} &
\colhead{$t_{\rm end}$} &
\colhead{$t_{\rm PT}$} & 
\colhead{Pd ($t_{\rm end}-t_{\rm tran}$)} &
\colhead{OPTd ($t_{\rm end}-t_{0}$)}
\\
\colhead{} & 
\colhead{(mag 100 day\textsuperscript{-1})} & 
\colhead{(mag 100 day\textsuperscript{-1})} & 
\colhead{(mag 100 day\textsuperscript{-1})} & 
\colhead{(mag)} & 
\colhead{(mag)} &
\colhead{(mag)} &
\colhead{(mag)} &
\colhead{(days)} &
\colhead{(days)} & 
\colhead{(days)} &
\colhead{(days)}
}
\startdata
ASASSN-14gm & 0.43 (0.36) & 0.43 (0.36) & 0.77 (0.02) & $-$16.87 (0.01) & $-$16.54 (0.07) & $-$14.47 (0.05) & - & 43.25 (10.36) & 110.93 (0.12) & - & 43.25 (10.36) \\
ASASSN-14jb & 0.25 (0.14) & $-$0.03 (0.14) & - & $-$14.93 (0.01) & - & - & 31.50 (19.7) & - & - & - & - \\
ASASSN-15bb & 0.60 (0.130) & 0.47 (0.06) & 2.65 (1.22) & $-$17.52 (0.01) & $-$17.27 (0.02) & $-$15.23 (0.02) & 45.75 (11.1) & 84.75 (0.0) & 104.82 (0.29) & 39.00 (11.1) & 84.75 (0.01) \\
ASASSN-15fz & 2.02 (0.46) & 0.94 (0.19) & - & $-$18.03 (0.02) & - & - & 39.75 (14.3) & - & - & - & - \\
ASASSN-15jp & 2.97 (0.07) & - & - & $-$18.216 (0.02) & - & - & - & - & - & - & - \\
LSQ13dpa & 0.28 (2.01) & 0.28 (2.01) & 0.68 (1.65) & $-$16.69 (0.01) & $-$16.33 (0.17) & $-$14.57 (0.17) & - & 115.00 (29.31) & 128.46 (0.47) &- & 115.00 (29.31) \\
LSQ15ok & 1.03 (0.66) & 1.03 (0.11) & - & $-$16.93 (0.01) & $-$15.55 (0.07) & - & - & 88.75 (8.05) & - & - & 88.75 (8.05) \\
SN~2012A & 1.74 (0.47) & 0.94 (0.09) & 1.04 (0.28) & $-$16.14 (0.01) & $-$15.39 (0.03) & $-$12.33 (0.02) & 45.47 (13.18) & 87.11 (25.67) & 109.87 (0.12) & 41.64 (15.84) & 87.11 (25.67) \\
SN~2012aw & 1.27 (0.07) & 1.27 (0.07) & - & $-$16.72 (0.01) & - & - & - & - & - & - & - \\
SN~2013ab & 1.33 (0.01) & 1.33 (0.01) & - & $-$17.05 (0.01) & - & - & - & - & - & - & - \\
SN~2013ai & 1.61 (0.01) & 1.61 (0.01) & - & $-$15.9 6(0.01) & - & - & - & - & - & - & - \\
SN~2013by & 2.80 (0.01) & 2.80 (0.01) & 1.20 (0.03) & $-$18.35 (0.01) & $-$17.01 (0.01) & $-$13.85 (0.01) & - & 21.75 (0.01) & 82.58 (0.03) & - & 21.75 (0.01) \\
SN~2013ej & 2.34 (0.79) & 2.34 (0.79) & 1.58 (0.08) & $-$17.74 (0.14) & $-$16.66 (2.31) & $-$12.43 (0.04) & - & 65.25 (24.92) & 95.8741 (0.277419) & - & 65.25 (24.92) \\
SN~2013gd & 1.25 (0.17) & 0.64 (0.02) & - & $-$16.38 (0.01) & $-$15.85 (0.03) & - & 32.75 (2.88) & 103.50 (0.01) & 111.93 (1.79) & 70.75 (2.88) & 103.50 (0.01) \\
SN~2013gu & 0.94 (0.17) & 1.91 (0.04) & - & $-$17.83 (0.01) & - & - & 16.50 (2.19) & - & - & - & - \\
SN~2013hj & 1.13 (0.13) & 1.59 (0.66) & 1.07 (0.08) & $-$17.86 (0.08) & $-$16.79 (0.06) & $-$14.48 (0.07) & 26.11 (18.74) & 82.50 (2.09) & 103.17 (2.71) & 56.39 (16.7) & 82.50 (2.09) \\
SN~2014cw & 1.35 (0.30) & 0.54 (0.09) & - & $-$16.75 (0.01) & - & - & 29.50 (5.50) & - & - & - & - \\
SN~2014dw & 2.37 (0.15) & 2.84 (0.07) & - & $-$17.01 (0.01) & - & - & 17.75 (3.83) & 68.50 (0.01) & - & 50.75 (3.83) & 68.50 (0.01) \\
\enddata
\end{deluxetable*}

\end{document}